\documentclass[journal]{IEEEtran}

\ifCLASSINFOpdf
\usepackage[pdftex]{graphicx}
\else
\fi
\usepackage{stfloats}
\usepackage{pbox}
\usepackage{cite}
\usepackage{graphicx}
\usepackage{balance}
\usepackage{float}
\usepackage{amsmath}
\usepackage[utf8]{inputenc}
\usepackage[final]{pdfpages}
\usepackage{pdfpages}
\usepackage{lipsum}
\usepackage{textcase}
\usepackage{amsmath,esint}
\usepackage{longtable}
\usepackage{array}
\usepackage{tabulary}
\usepackage[section]{placeins} 
\usepackage[normalem]{ulem}
\usepackage{array,multirow}
\usepackage{amssymb}
\usepackage{subcaption}
\usepackage{booktabs}
\usepackage{makecell}
\PassOptionsToPackage{hyphens}{url}\usepackage{hyperref}
\RequirePackage[hyphens]{url}

\newcolumntype{P}[1]{>{\centering\arraybackslash}p{#1}}

\begin{document}

\title{Multiple Ray Received Power Modeling for mmWave Indoor and Outdoor Scenarios}

\author{Wahab Khawaja, Ozgur Ozdemir, Fatih Erden, Ender Ozturk and Ismail Guvenc

\thanks{This work has been supported in part by NASA under the Federal Award ID number NNX17AJ94A.} 

\thanks{All the authors are with the Department of Electrical and Computer Engineering, North Carolina State University, Raleigh, NC 27606 (e-mail:\{wkhawaj, oozdemi, ferden, eozturk2, iguvenc\}@ncsu.edu).}
}

\maketitle

\begin{abstract}
Millimeter-wave (mmWave) frequency bands are expected to be used for future 5G networks due to the availability of large unused spectrum. However, the attenuation at mmWave frequencies is high. To resolve this issue, the utilization of high gain antennas and beamforming mechanisms are widely investigated in the literature. In this work, we considered mmWave end-to-end propagation modeled by individual ray sources, and explored the effects of the number of rays in the model and radiation patterns of the deployed antennas on the received power. It is shown that taking the dominant two rays is sufficient to model the channel for outdoor open areas as opposed to the indoor corridor which needs five dominant rays to have a good fit for the measurement and simulation results. It is observed that the radiation pattern of the antenna affects the slope of the path loss. Multi-path components increase the received power, thus, for indoor corridor scenario, path loss according to the link distance is smaller for lower gain antennas due to increased reception of reflected components. For an outdoor open area, the slope of the path loss is found to be very close to that of the free space.


\end{abstract}

\section{Introduction} \label{Section:Introduction}
There has been a significant increase in the number of smart communication devices and high data rate applications in the last decade. This trend is expected to grow rapidly in the future~\cite{smart_phone}. However, the available spectrum at the sub-6~GHz band is limited. Higher frequency bands (e.g. mmWave bands) are not heavily utilized, thus, offer larger bandwidths for wireless communication systems. Therefore, research efforts have been concentrated on exploring higher frequencies as an alternative to the sub-6~GHz band. The opening of the mmWave spectrum for mobile usage by FCC~\cite{FCC_28G} has given a boost to the current researches to best utilize these bands. However, mmWave communication suffers from its inherent high free space attenuation as well as high penetration losses.

\begin{figure}[!t]
\centerline{\includegraphics[width=0.87\columnwidth]{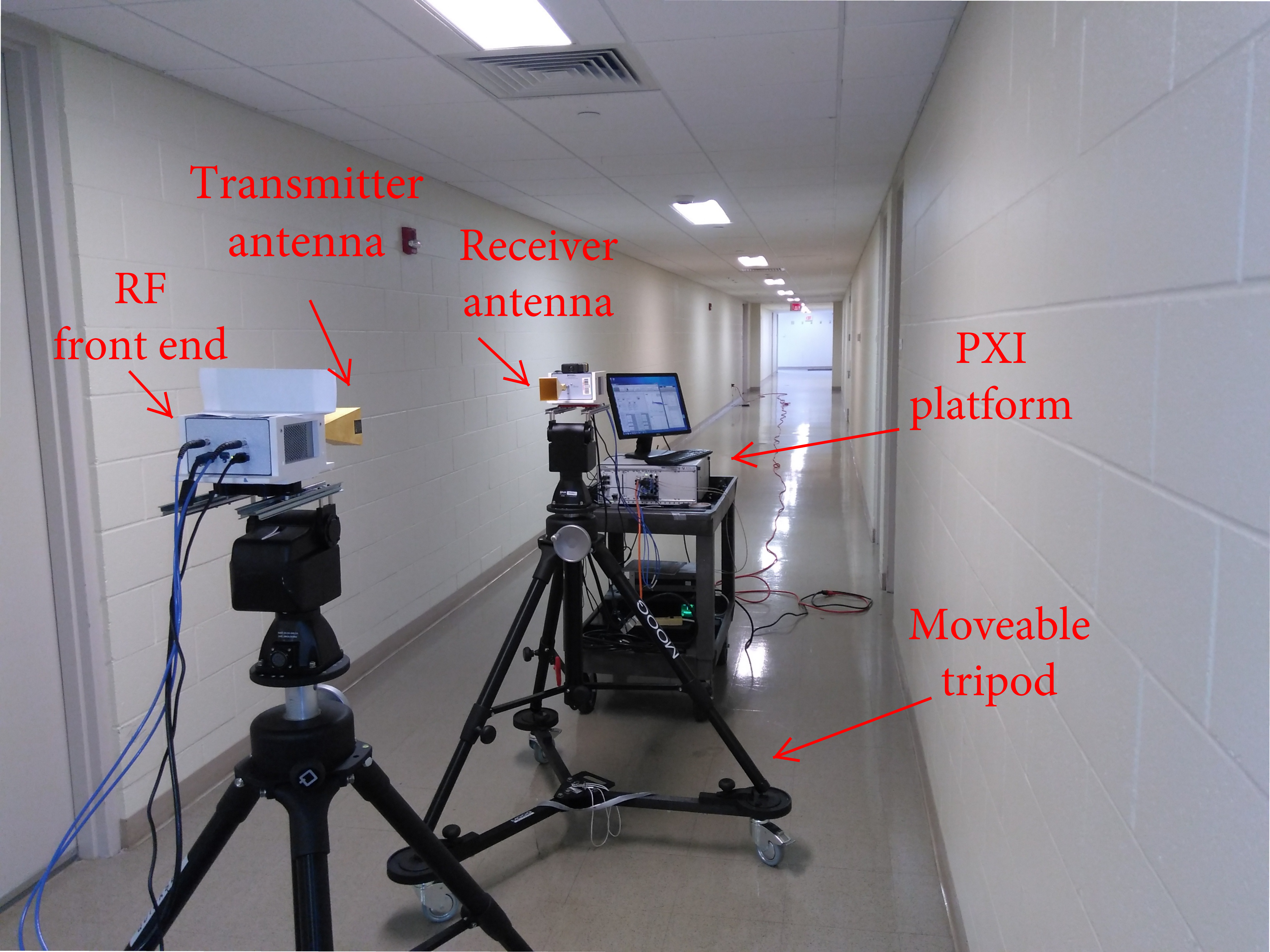}}
\caption{Indoor corridor propagation setup at the basement of Engineering Building II, North Carolina State University. } \label{Fig:corridor_scenario}
\end{figure}

\begin{figure}[!t]
\centering\vspace{-3mm}
\centerline{\includegraphics[width=0.87\columnwidth]{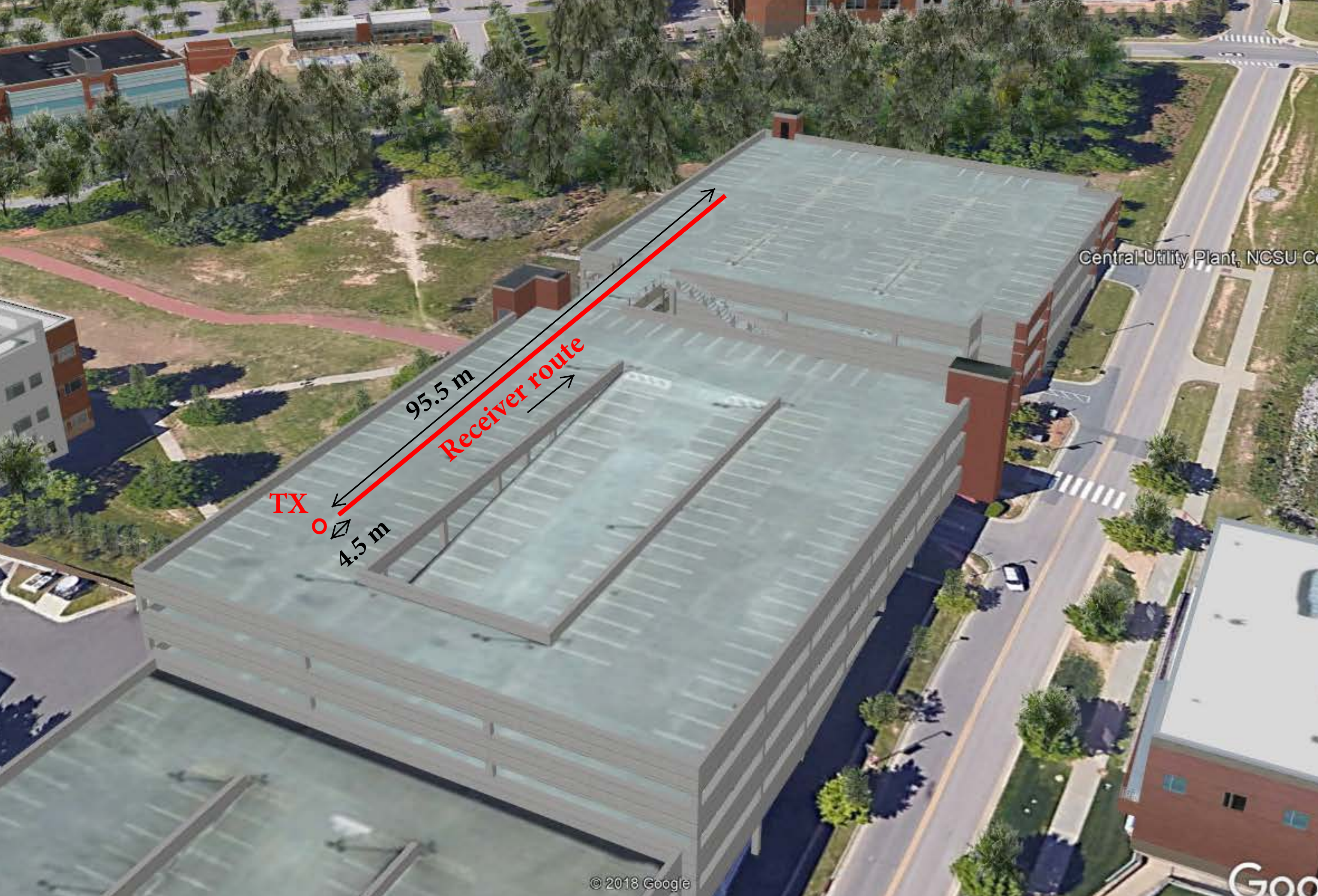}}
\caption{Outdoor measurement setup at the top floor of a multi-story car park, Centennial Campus, North Carolina State University. } \label{Fig:outdoor_scenario}
\end{figure}

In this work, we used measurements, analytical ray modeling and ray tracing simulations to model LoS characteristics of a mmWave communication channel in a corridor type indoor and open space outdoor environments at 28~GHz frequency band. We analytically calculated received signal properties using the dominant five-ray and two-ray received power model based on first-order reflections for the indoor corridor and outdoor open area, respectively. To compare with our analytical results, measurements were conducted at North Carolina State University using a PXI-based channel sounder platform from National Instruments, and two sets of directional horn antennas with gains $17$~dBi and $23$~dBi at $28$~GHz. The test setup used indoor and outdoor are shown in Fig.~\ref{Fig:corridor_scenario} and Fig.~\ref{Fig:outdoor_scenario}. 

The rest of the paper is organized as follows. Readers will find a comprehensive literature review as well as a summary of our contributions in Section~\ref{Section:Literature Review}. Section~\ref{Section:RX_pwr_model} includes details on received power modeling for indoor and outdoor environments. Section~\ref{Section:Exp_ray_tracing_setup} covers experimental and ray tracing simulations setup. In Section~\ref{Section:Num_rays_pow}, number of rays and percentage power sum of dominant five rays with total power of rays is provided. In Section~\ref{Section:RX_power_PL_analysis}, results of measurements, simulations and calculations for received power are given. In Section~\ref{Section:Five_two_compare}, a detailed discussion is presented for five ray and two ray models. Section~\ref{Section:K_factor} provides Ricean $K$-factor analysis and the paper ends with concluding remarks in Section~\ref{Section:Conclusions}.

\begin{table*}[!t]
\caption{Related work in the literature on mmWave channel modelling using ray tracing. \label{Table:Literature_comparison}}
\begin{tabular*}
{\textwidth}{@{\extracolsep{\fill}}lcccc}\toprule
Literature & Number of rays & Frequency &Maximum distance & Reported channel statistics \\
\midrule
\cite{two_ray1} &Two&sub-6~GHz and mmWave&10 km&Received power, two ray model,\\&&&& 
break point,distance based on first \\&&&& Fresnel zone \\
\cite{two_ray2}&One,two,five,twenty&100 MHz, 1800~MHz, 2400~MHz&10~km&Path loss, two ray model,\\ &&&&effect of first Fresnel zone on \\&&&& path loss exponent  \\
\cite{two_ray3}&Two&1.5~GHz&1 km&Two ray model, path loss exponent for \\&&&&vertically and horizontally \\&&&&polarized signals \\
\cite{three_ray1}&Three&3.6~GHz, 10.6~GHz&100 m&Path loss, three ray model for UWB\\&&&& propagation \\
\cite{three_ray3}&Three&1900~MHz&400~m&Three ray propagation model for \\&&&& PCS and $\mu$-cellular services \\
\cite{multi_ray1}&Sixty two&0.06~THz-1~THz&6 m&Distance and frequency selective \\&&&&characteristics, coherence bandwidth,\\&&&& channel capacity, and temporal broadening \\&&&& analysis\\
\cite{multi_ray2}&Nine&60 GHz&60~m LOS, 25~m NLOS&Received power, indoor corridor power\\&&&& distribution comparison with \\&&&& Rayleigh and Rician \\
\cite{multi_ray3}&Two, four, five&2.4~GHz&50~m&Received power analysis in open \\&&&& and closed corridors \\
\cite{multi_ray5}&Two, multiple rays&94~GHz&6 m&Path loss, multipath analysis \\
\cite{multi_ray6}&Two, four&94`GHz&1.5~m&Received power, multipath analysis for\\&&&& radars \\
\cite{multiple_ray3}&Two, four, six, ten&2.4~GHz&10~m&Path loss \\
This study&Two, five&28~GHz&40 m indoor, 100 m outdoor&Received power, path loss, effect of\\&&&&antenna gain, adequacy analysis on number\\&&&&of rays using z-test and Ricean K-factor\\
\bottomrule
\end{tabular*} 
\end{table*}

\section{Literature Review and Contributions} \label{Section:Literature Review}
Various approaches have been proposed in the literature to overcome the high attenuation problem at mmWave frequencies~\cite{wahab_refl_journal, indoor_wahab}. A common method is to increase the gain or directivity of the antennas~\cite{ant1,ant2}. High directivity is obtained either by beamforming or deploying directional antennas (e.g. horn antennas). In addition to antenna type, material characteristics of the objects in the environment also play an important role in figuring the propagation statistics~\cite{refl1,refl2}. One way of modeling propagation statistics is using ray tracing. In the literature, different types of indoor geometries either in line-of-sight~(LoS) or non-line-of-sight~(NLoS) scenarios for a wide variety of frequency bands are investigated using ray tracing software~\cite{Mrugesan_2012, Karstensen_2016, Helhel_2006, AlAbdullah_2017}. 

In this work, we modeled the end-to-end propagation as individual ray sources. For the indoor environment, five rays are used in calculations. One is the LoS and four are the reflected rays from two walls, ceiling, and ground. Each ray source contributes to the resulting received power. Contributions of the reflected rays are found to increase with the link distance. This is because when transmitter and receiver antennas are close, reflected fields are rejected by the receiver antenna because of its directional pattern. Together with high Fresnel reflection coefficient values along with the link, we observe an increase in the received power compared to free space i.e. the slope of the path loss is smaller than that of the free space for indoor. For outdoor open area, two ray model is found to be sufficient to model the received power and because of the absence of three first-order reflections, no obvious difference between path loss slopes have been detected. The analytical modeling results based on ray sources are compared with measurement and ray tracing simulation results.

We also made a comparative analysis of the measurements with five ray and two ray analytical models and ray tracing simulations with five rays are provided for the indoor environment. The comparative analysis is carried out using $z$-test of the path loss model parameters. The $z$-test values indicate that the two ray model does not provide a close match to the measurement path loss for the indoor corridor. On the other hand, five ray model provides a close fit to the measurement data.

The ratio of power sum of dominant five rays to power sum of total rays obtained from measurements is also provided in this work. The percentage is greater than 90\% for all the scenarios, which indicates that five rays are sufficient for modeling. The Ricean $K$-factor is also provided to study the contribution of LoS ray and diffuse rays over the link for two different gain antennas.
Table~\ref{Table:Literature_comparison} shows the related work in the literature, where ray tracing is used. Comparison of the available literature with our work highlights the following distinctions of our work:

\begin{itemize}
\item Propagation modeling based on dominant rays at 28~GHz is considered in our work.
\item Five dominant rays were found to be adequate for indoor corridor propagation modeling whereas, two dominant rays were found to adequately model the open area outdoors. The antenna gain of each individual ray is modeled based on its geometric position from the radiation pattern of the antenna provided in the datasheet.
\item Resolvable distance of the rays compared to the LoS as a function of the link distance are also provided. Smaller than this resolvable distance, the rays will be superimposed coherently with the LoS component.
\item A polarization dependent reflection coefficient for different materials is used at 28~GHz.
\item A $z$-test is also performed for comparison of parameters of two ray and five ray path loss models obtained analytically, through ray tracing simulations and measurements.
\item A commonly occurring scenario for future 5G deployments are closely positioned transmitter and receivers at indoor corridors. This commonly occurring scenario in a typical indoor corridor environment is studied.
\end{itemize}

There are other works in the literature in which five, even more, first-order reflections are taken into account~\cite{Li_2016,multi_ray2,Zeyde_2018,multiple_ray1,multiple_ray2,multiple_ray3, multi_ray1,multi_ray3, multi_ray5, multi_ray6}. We did not consider the higher number of rays~(considering higher-order reflections). This is because most of the received power comes from the LoS signal and first-order reflections. Moreover, considering higher-order reflections increases the complexity of the model unnecessarily compared to their contribution to the received power. Therefore, our model based on LoS and first-order reflections provide a robust and simple way to calculate received power in corridors and similar shaped indoor environments. Similarly, for outdoor open area two ray model is sufficient to model the received power.

\section{Received Power Modeling Based on Dominant Rays for Indoor Corridor and Outdoor Open Area}~ \label{Section:RX_pwr_model}
In this section, we will first discuss antenna radiation pattern effects on propagation. Later, a received power calculation model based on dominant LoS signal and reflected rays in the indoor corridor~(five ray model) is presented. Two ray model, as a special case of five ray model, is used for outdoor open area.

\subsection{Antenna Radiation Pattern and Propagation Effects}  \label{Section:Antenna_rad_pattern}
The antenna radiation pattern plays an important role in modeling the propagation characteristics of directional mmWave links. In the model, we used two directional horn antenna sets which have different gains and respective half-power beamwidths~(HPBWs) in the azimuth and elevation planes. We represent the 3D antenna gain as a surface area extended on a sphere at a distance $d$ with a given solid angle~$\Omega$. The surface area $A$ subtended by the antenna gain at a distance $d$ from the source is $A = d^2\Omega$, where the solid angle $\Omega$ is given as:
\begin{align}
    \Omega = \int_{\phi=0}^{2\pi}\int_{\theta=0}^{\pi} \frac{P_{\rm rad}(d,\theta,\phi)}{P_{\rm max}} d\delta,~~
    d\delta = \sin\theta d\theta d\phi,
\end{align}
where $P_{\rm rad}(d,\theta,\phi)$ is the radiated power from the antenna in spherical coordinates as a function of distance $d$, elevation and azimuth angles of $\theta$ and $\phi$, respectively. $P_{\rm max}$ is the maximum radiated power. 
The propagation from the transmitting antenna is modeled as a spherical wavefront. The majority of the radiated power is concentrated over the area covered by the solid angle represented by $\Delta \theta$ and $\Delta \phi$, where these two angles represent the antenna HPBWs in the elevation and azimuth planes, respectively. Moreover, if $\Delta \theta$ and $\Delta \phi$ are small, we can approximate the area extended by $\Delta \theta$ and $\Delta \phi$ in space as $A_{\rm hpa}=d_{\rm f}^2\Delta \theta\Delta \phi$ at a fixed distance $d_{\rm f}$ in the far-field region. The rays lying in this region will have significantly higher gain compared to the rays lying outside this area. 

\begin{figure}[!t]
\centering
\centerline{\includegraphics[width=\columnwidth]{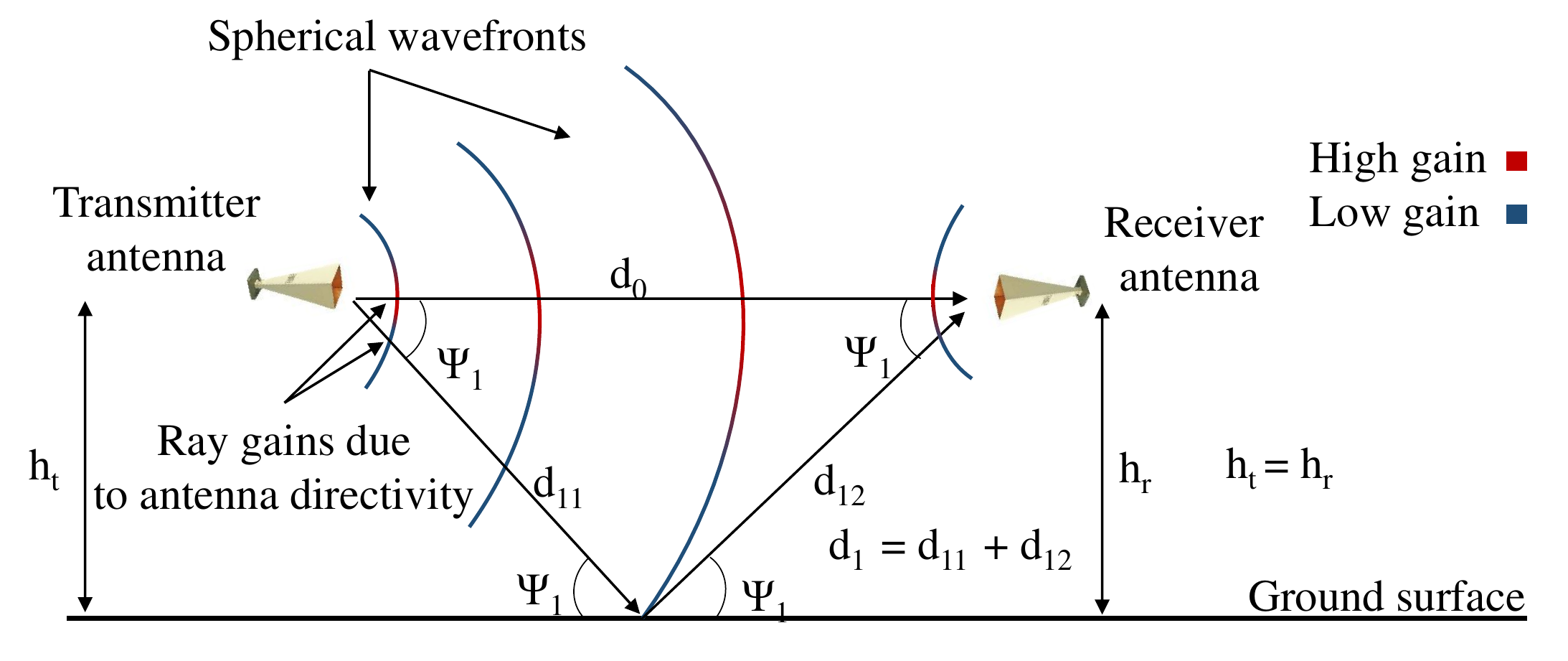}}
\caption{Propagation of line-of-sight and ground reflected component from transmitter antenna towards receiver antenna when their heights~($h_{\rm t}$ and $h_{\rm r}$) are the same.} \label{Fig:grazing_angle}
\end{figure}

\subsection{Ray Resolution Along the Link Distance} \label{Section:Ray_resol}
Our channel sounder setup can resolve any two rays at a spatial distance represented as $d_{\rm {res}} = 0.585~\rm{m}$. Consider the case of two ray modeling for a given height of the transmitter and receiver represented as $h_{\rm t}$ and $h_{\rm r}$, respectively (Fig.~\ref{Fig:grazing_angle}). When the link distance $d$ between the transmitter and receiver is increased such that the difference between the paths traveled by any two rays is smaller than $d_{\rm res}$, those rays cannot be resolved, thus can be measured as a superposition. The relevant inequality is as follows:   
\begin{equation}
    \sqrt{(h_{\rm t} + h_{\rm r})^2 + d^2}- d > d_{\rm res}~. \label{Eq:Ray_resolvable} 
\end{equation}
\begin{figure}[!t]
\centering
\centerline{\includegraphics[width=\columnwidth]{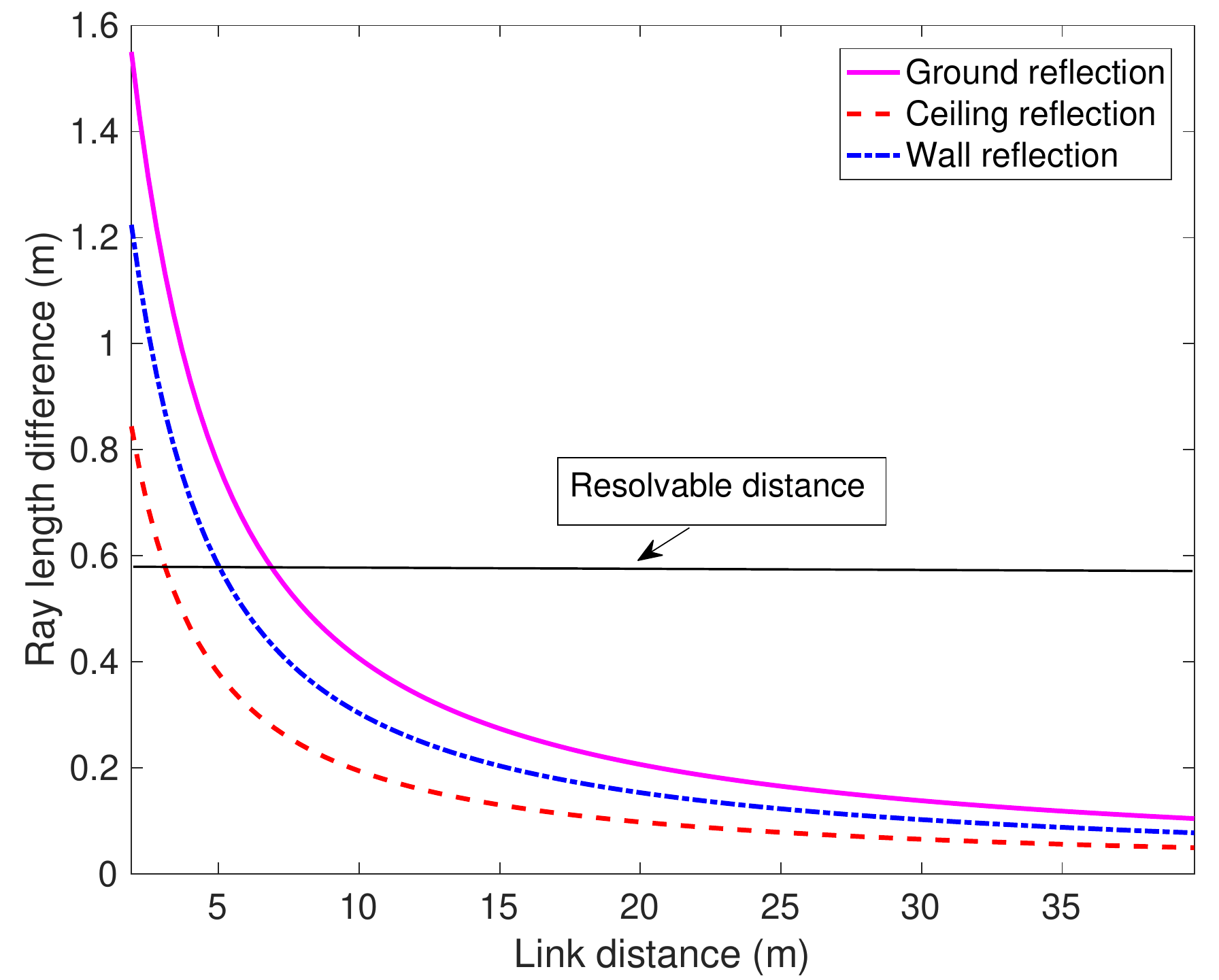}}
\caption{The difference of ray lengths with the LoS, plotted as a function of link distance.} \label{Fig:Ray_distances}
\end{figure}
Similarly, for the indoor corridor, the rays reflected from the ground, ceiling, and walls may not be resolvable depending on the link distance $d$. Fig.~\ref{Fig:Ray_distances} shows the difference of path distances of the rays reflected from ground, ceiling, and walls with respect to the LoS ray. In Fig.~\ref{Fig:Ray_distances}, the reflected rays are considered to be independent of each other. According to Fig.~\ref{Fig:Ray_distances}, the ray from the ceiling is the first to get unresolved at $3.1$~m compared to ground reflected ray, which gets unresolved at $7$~m. The rays from the two walls are not resolvable after $5$~m. This indicates that the path of the reflected ray from the ceiling is the smallest compared to the paths of the remaining three rays.

\subsection{Received Power Modeling for Indoor Corridor} \label{Section:Rec_pwr_five_ray}
The received signal is given by $R(n) = S(n)\circledast H(n)$, where $S(n)$ represents the transmitted signal, $H(n)$ is the impulse response of the channel and $\circledast$ is the convolution operation. In case that received and transmitted signals are known, channel impulse response~(CIR) could be obtained by applying deconvolution.


\begin{figure}[!t]
\centering
\centerline{\includegraphics[width=\columnwidth]{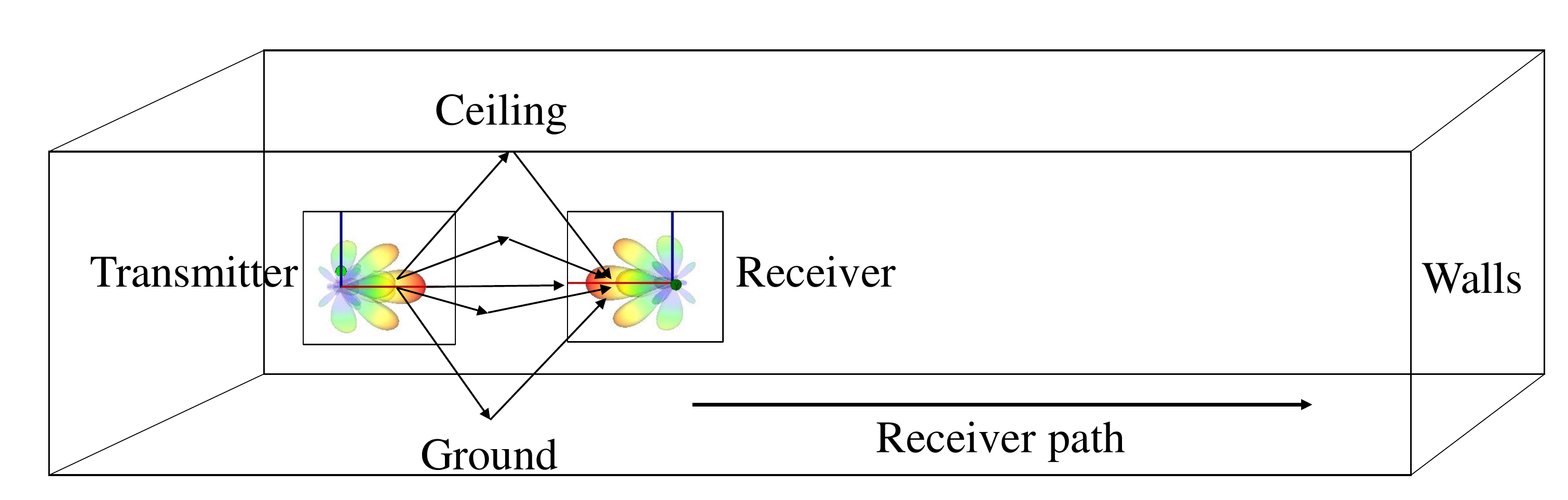}}
\caption{The layout of the indoor corridor propagation environment. } \label{Fig:scenario_layout}
\end{figure}

In this work, we considered the CIR in the indoor corridor~(similar to rectangular waveguide) and outdoor open area. The height of the transmitter and receiver are kept the same throughout the experiments. The indoor corridor propagation layout is shown in Fig.~\ref{Fig:scenario_layout}. For the corridor scheme, there are five dominant rays at any given distance from the transmitter towards the receiver. One is the LoS and the four others are the reflected rays from the ground, ceiling and two walls. The strength of the reflected rays is dependent mainly on the antenna radiation pattern.

As the distance of the receiver increases from the transmitter moving in a straight line received power coming from reflected rays increase as well. Due to the geometry of the test setup, as the link distance increases, reflected rays gets closer to the boresight of the received antenna, thus, captured with a higher gain. As a result of this, the difference between power value calculated taking only free space path loss into account and the five ray received power increases in favor of five ray model. 

The contribution of the reflected rays to the overall received power is also dependent on the Fresnel reflection coefficients. Reflected rays of more than first-order have a significantly smaller contribution to the received power compared to first-order reflections. Therefore, in our model, we can safely ignore their contributions. Let $R_0$ represent the received LoS component given as:
\begin{align}
    R_0(n) =  &\bigg[\frac{\lambda}{4\pi d_0}\sqrt{G_{\rm T}(\theta^{\rm (TX)},\phi^{\rm (TX)}) G_{\rm R}(\theta^{\rm (RX)},\phi^{\rm (RX)})}\nonumber \\& S(n-\tau_0)\exp \bigg(\frac{-j2\pi d_0}{\lambda}\bigg) \Big|\pmb{\rho}^{\rm (TX)}_{0}\cdot\pmb{\rho}^{\rm (RX)}_{0}\Big| \bigg], \label{Eq:Rec_pwr_LOS}
\end{align} 
where $G_{\rm T}(\theta^{\rm (TX)},\phi^{\rm (TX)})$ is the gain of the antenna for the transmitter at elevation and azimuth angles of $\theta^{\rm (TX)}$ and $\phi^{\rm (TX)}$, respectively.  Similarly, $G_{\rm R}(\theta^{\rm (RX)},\phi^{\rm (RX)})$ is the gain of the antenna for the receiver at elevation and azimuth angles of $\theta^{\rm (RX)}$ and $\phi^{\rm (RX)}$, respectively, $\tau_0$ represents the delay of the LoS component given by $\tau_0 = \frac{d}{c}$, where $c$ is the speed of the light and $d$ is the distance of the LoS component, $\exp \big(\frac{-j2\pi d_0}{\lambda}\big)$ represents the phase of the LoS component, $\pmb{\rho}^{\rm (TX)}\cdot\pmb{\rho}^{\rm (RX)}$ represents the dot product between the polarization unit vectors of the electric field at the transmitter and receiver, respectively.

The gain of the antenna for the LoS ray in the azimuth and elevation planes at the transmitter and receiver is given as follows~\cite{Wireless_Insite}:

\begin{align}
    &\sqrt{G_{\rm T}\big(\theta^{\rm (TX)},\phi^{\rm (TX)}\big)G_{\rm R}\big(\theta^{\rm (RX)},\phi^{\rm (RX)}\big)} = \nonumber \\
    &\qquad g^{(\rm {TX},\theta)}\big(\theta^{\rm (TX)}, \phi^{\rm (TX)}\big)g^{(\rm {RX},\theta)}\big(\theta^{\rm (RX)}, \phi^{\rm (RX)}\big) \nonumber\\
    &\quad + g^{(\rm {TX},\phi)}\big(\theta^{\rm (TX)}, \phi^{\rm (TX)}\big)g^{(\rm {RX},\phi)}\big(\theta^{\rm (RX)}, \phi^{\rm (RX)}\big), \label{total_gain}
\end{align}
where $g^{(\rm {TX},\theta)}\big(\theta^{\rm (TX)}, \phi^{\rm (TX)}\big)$ and $g^{(\rm {TX},\phi)}\big(\theta^{\rm (TX)}, \phi^{\rm (TX)}\big)$ represents the direction of departure~(DoD) in the elevation and azimuth planes, respectively. Similarly, the direction of arrival~(DoA) in the elevation and azimuth planes are given as $g^{(\rm {RX},\theta)}\big(\theta^{\rm (RX)}, \phi^{\rm (RX)}\big)$ and $g^{(\rm {RX},\phi)}\big(\theta^{\rm (RX)}, \phi^{\rm (RX)}\big)$.

$g^{(\theta)}(\theta,\phi)$ can be expressed as follows:
\begin{equation}
    g^{(\theta)}(\theta,\phi) = \sqrt{|G_{\theta}(\theta, \phi)|} \exp(j\varphi_{\theta})
\end{equation}
where $G_{\theta}$ is the antenna gain and $\varphi_{\theta}$ is the relative phase of the $\theta$ component of a ray. If both the transmitter and receiver are aligned to their boresight, then the total gain given in \eqref{total_gain} is maximized. 

Similar to the LoS component, the four dominant received rays reflected from the environment , with the ray index $i=1,2,3,4$, is expressed as:
\begin{align}
        R_i(n)=&\bigg[\frac{\lambda \Gamma_i(\Psi_i)}{4\pi d_i}\sqrt{G_{\rm T}\big( \Psi_i^{(\rm Az)},\Psi_i^{(\rm El)} \big) G_{\rm R}\big(\Psi_i^{(\rm Az)},\Psi_i^{(\rm El)}  \big)} \nonumber \\& S(n-\tau_i) \exp \bigg(\frac{-j2\pi d_i}{\lambda}\bigg) \Big|\pmb{\rho}^{\rm (TX)}_{i}\cdot\pmb{\rho}^{\rm (RX)}_{i}\Big| \bigg], \label{Eq:Five_ray_comp}
\end{align}

The reflection coefficient $\Gamma_i(\Psi_i)$ also called Fresnel reflection coefficient for the relative permittivity of the ground material $\epsilon_r$ is given as:
\begin{equation}
    \Gamma_i(\Psi_i) = \frac{\sin\Psi_i - Y}{\sin\Psi_i + Y},
\end{equation}
where the value of $Y$ depends on the polarization and are given for vertical and horizontal polarization as follows:
\begin{equation}
     Y_{\rm v} = \frac{\sqrt{\epsilon_r - \cos^2\Psi_i}}{\epsilon_r}, ~~ Y_{\rm h} = \sqrt{\epsilon_r - \cos^2\Psi_i} \label{Eq:Permittivity}.
\end{equation}

If the link distance $d \to \infty$, then $\Psi \to 0$ and the gain of the reflected ray approaches to the LoS component gain and the Fresnel reflection coefficient, $\Gamma \to -1$.   

Let $E$ represent the average over time, and $P_{\rm R}$ represent the total received power, then $P_{\rm R}$, the coherent addition of the LoS and the reflected rays for $i=0,1,2,3,4$, is given as:
  \begin{equation}
 \begin{aligned}
  P_{\rm R}(d_i)={\rm E}\bigg[\bigg|R_0(n) + \sum_{i=1}^{4}R_i(n)\bigg|^2 \bigg]. \label{Eq:Total_power} 
 \end{aligned}
 \end{equation}
Equation (\ref{Eq:Total_power}) can be rewritten for $d$ values such that the reflected rays can be resolvable~(see Section~\ref{Section:Ray_resol}, Fig.~\ref{Fig:Ray_distances}) from each other: 
  \begin{equation}
 \begin{aligned}
  P_{\rm R}(d_i)={\rm E}\big[\big|R_0(n)\big|^2\big] + \sum_{i=1}^{4}{\rm E}\big[\big|R_i(n)\big|^2\big]. \label{Eq:Total_power_noncoh} 
 \end{aligned}
 \end{equation} 

From (\ref{Eq:Rec_pwr_LOS}), (\ref{Eq:Five_ray_comp}), if $S(n) \approx S(n-\tau_0) \approx S(n-\tau_i)$, and $P_{\rm T} = \rm{E}\big[ |S(n)|^2\big]$ where $P_{\rm T}$ is the transmitted power. Moreover, for the LoS component, the XPD (cross polarization discrimination) factor is negligible for vertical-vertical~(VV) and horizontal-horizontal~(HH) antenna orientations. Similarly, for the reflected rays, the diffuse scattering is small due to smooth reflecting surfaces leading to small XPDs. Therefore, the dot product of the polarization vectors $|\pmb{\rho}^{\rm (TX)}_{i}\cdot\pmb{\rho}^{\rm (RX)}_{i}\Big|$ can be taken as 1 for the LoS and reflected rays. Therefore, the total received power from (\ref{Eq:Total_power}) can be written as follows:

\begin{align}
           P_{\rm R}(d_i)=&P_{\rm T}\bigg(\frac{\lambda}{4\pi}\bigg)^2\Bigg|\frac{\sqrt{G_{\rm T}(\theta^{\rm (TX)},\phi^{\rm (TX)})G_{\rm R}(\theta^{\rm (RX)},\phi^{\rm (RX)})}}{d_0} + \nonumber \\& \sum_{i=1}^{4} \Gamma_i(\Psi_i)\sqrt {G_{\rm T}\big(\Psi_i^{(\rm Az)},\Psi_i^{(\rm El)} \big)G_{\rm R} \big(\Psi_i^{(\rm Az)},\Psi_i^{(\rm El)} \big)} \nonumber \\& \frac{\exp(-j\Delta \Omega_i)}{d_i}\Bigg|^2, \label{Eq:Total_power_five_ray}
\end{align}
where $\Delta \Omega_i = \frac{2\pi(d_i-d_0)}{\lambda}$ for $i=1,2,3,4$. Additionally, if the heights of the antennas are not the same and/or not aligned to the boresight, we have additional attenuation due to smaller antenna gain. This attenuation will decrease with the increase in distance between the transmitter and the receiver. 

Considering the $i^{\rm th}$ individual reflected ray at a given link distance, we can write the received power as follows:
\begin{align}
           P_{\rm R}(d_i)=&P_{\rm T}\bigg(\frac{\lambda}{4\pi d_i}\bigg)^2\Gamma_i^2(\Psi_i)G_{\rm T}\big(\Psi_i^{(\rm Az)},\Psi_i^{(\rm El)} \big) \nonumber\\& G_{\rm R} \big(\Psi_i^{(\rm Az)},\Psi_i^{(\rm El)} \big) \label{Eq:Ind_ray}. 
\end{align}
From~(\ref{Eq:Ind_ray}), it can be observed that the received power of the $i^{\rm th}$ reflected ray approaches to the LoS ray at distance $d_i$ when, 1) the antenna gains at the transmitter and receiver side are equal to the boresight antenna gains, 2) the reflection coefficient is 1.

\subsection{Received Power Modeling for Outdoor Open Area}
The two ray model can be considered as a special case of the five ray model. The two ray model is used for received power modeling in outdoor open area assuming that antenna heights are significantly high. The contribution of any other rays from far off scatterers is small for the open area and is ignored. In the two ray modeling, the received power is dependent on the LoS and ground reflected component~(GRC). Therefore, the total received power is given as follows:
\begin{align}
   P_{\rm R}(d_0,d_1) = &P_{\rm T}\bigg(\frac{\lambda}{4\pi}\bigg)^2\Bigg|\frac{\sqrt{G_{\rm T}(\theta^{\rm (TX)},\phi^{\rm (TX)})G_{\rm R}(\theta^{\rm (RX)},\phi^{\rm (RX)})}}{d_0} \nonumber \\& +  \Gamma_1(\Psi_1)\sqrt{G_{\rm T}\big(\Psi_1^{(\rm Az)},\Psi_1^{(\rm El)} \big) G_{\rm R}\big(\Psi_1^{(\rm Az)},\Psi_1^{(\rm El)}\big)} \nonumber \\& \frac{\exp(-j\Delta \Omega)}{d_1}\Bigg|^2, \label{Eq:Total_power_coh}
\end{align}
where $\Delta \Omega = \frac{2\pi(d_1 - d)}{\lambda}$ is the phase difference between the LoS and the GRC signals.

\subsection{Polarization Effects on the Received Power} \label{Section:XPD}
The polarization of electric fields should be taken into account. There are two co-polarized configurations based on antenna orientation used in the measurements, namely VV and HH. The difference in VV and HH antenna orientations is subject to the antenna radiation pattern in the azimuth and elevation planes. However, even though the whole patterns are different in two orthogonal planes, as the HPBWs are the same for both horn antenna sets, no significant difference in the antenna radiation patterns has been observed due to antenna orientation.

Cross polarization of vertical-horizontal~(VH) is also introduced to study the XPD factor in the indoor corridor. Considering the channel stationary, we can obtain the XPD factor between the transmitter and receiver as follows:
\begin{equation}
    \rho  = \rm{E}\Bigg(\frac{P_R^{\rm{(VV)}}(d)}{P_R^{\rm{(VH)}}(d)}\Bigg)~~\rm{or}~~ \rho  = \rm{E}\Bigg(\frac{P_R^{\rm{(HH)}}(d)}{P_R^{\rm{(VH)}}(d)}\Bigg),  
\end{equation}
where $P_R^{\rm{(VV)}}$, $P_R^{\rm{(VH)}}$ and $P_R^{\rm{(HH)}}$ are the received powers for VV, VH and HH antenna orientations, respectively, and E$(\mathord{\cdot})$ denotes the expected value. A major use of XPD factor is that it helps to study the interaction of the antennas of different beamwidths with the surroundings when cross polarization is not negligible.  

\subsection{Path Loss Modeling} \label{Section:Path_loss}
The path loss obtained from the received power measured at different distances from the transmitter is given as follows:
\begin{align}
    L(d)~[\rm{dB}] = 10\log_{10}\bigg(\frac{P_T}{P_R(d)}\bigg).
\end{align}
An alpha-beta model for the path loss modeling~\cite{winner} is given as: 
\begin{align}
    L(d)~\rm{[dB]} = \beta + 10\alpha \log_{10}(d) + X, \label{Eq:PL}
\end{align}
where $\beta$ is the y-intercept in dB, $\alpha$ is the slope and $X$ is a random variable and $X \sim \mathcal{N}(0,\,\sigma^{2})$, where $\sigma^{2}$ expressed in dB is the variance of $X$. A least square regression is used to fit a regression line~(best fit) to the data.

\section{Experimental and Ray Tracing Simulations Setup} \label{Section:Exp_ray_tracing_setup}

In this section, an indoor and outdoor experimental setup, as well as the ray tracing simulation setup, are discussed. 

\subsection{Indoor and Outdoor Measurement Setup}
Indoor corridor measurements were carried out at the basement of the Engineering Building II, North Carolina State University, shown in Fig.~\ref{Fig:corridor_scenario}. The walls in the corridor are 3 layered drywall, the ceiling is Armstrong type ceiling and the ground is a concrete grinded surface. The measurements were carried out using NI mmWave transceiver system operating at $28$~GHz. The description of the NI mmWave transceiver system is provided in~\cite{outdoor_wahab}. Two horn antenna sets with gains $17$~dBi and $23$~dBi were used in the measurements. The HPBWs of $17$~dBi antennas are $26^{\circ}$ and $24^{\circ}$ in the E and H planes, respectively. The HPBWs for the $23$~dBi antennas in the E and H planes are $9.6^{\circ}$ and $11^{\circ}$, respectively. 

The height of the transmitter and receiver from the ground was fixed to $1.44$~m, whereas, the distance of the transmitter and receiver from the ceiling was $0.9$~m. The distance from either of the walls to the antennas was $1.24$~m. The transmitter was kept at a fixed position, whereas the receiver was moved in a straight line away from the transmitter at constant intervals of $0.3$~m starting from $1.9$~m to $39.7$~m. Laser alignment is used between the transmitter and the receiver at every step.

The outdoor measurements were carried out at the top floor of a multi-story car park at North Carolina State University shown in Fig.~\ref{Fig:outdoor_scenario}. Similar to the indoor corridor measurements, the transmitter was kept at a fixed place, and the receiver was moved in steps of $5$~m beginning from $4.6$~m to $100$~m. The height of the transmitter and receiver was $1.09$~m. For both indoor and outdoor measurements, the transmit power has been set to $0$~dBm.    

\begin{figure}[!t]
\centering
\centerline{\includegraphics[width=\columnwidth]{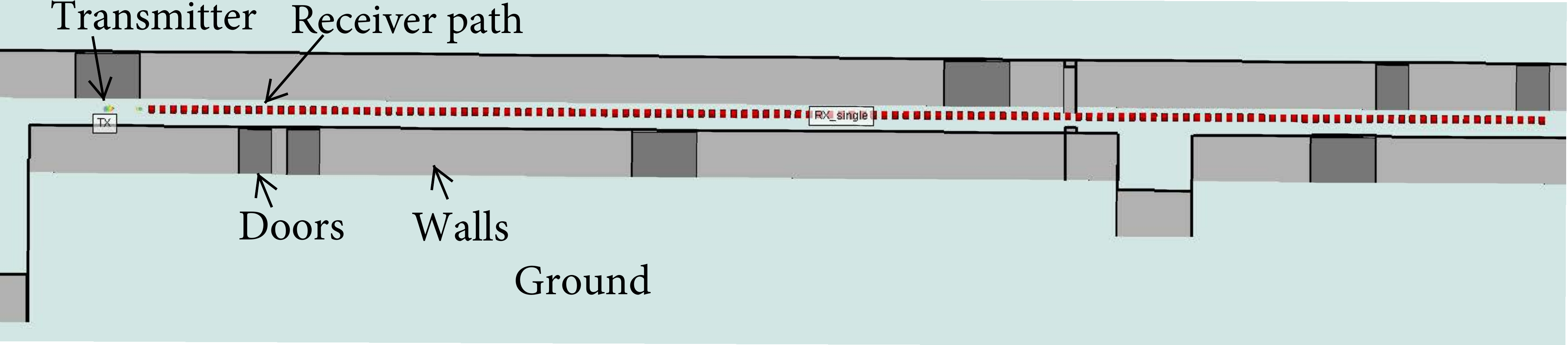}}
\caption{Indoor corridor scenario created in Wireless InSite for ray tracing simulations. } \label{Fig:ray_tracing_scenario}
\end{figure}

\begin{figure}[t] 
    \centering
	\begin{subfigure}{\columnwidth}
    \centering
	\includegraphics[width=\textwidth]{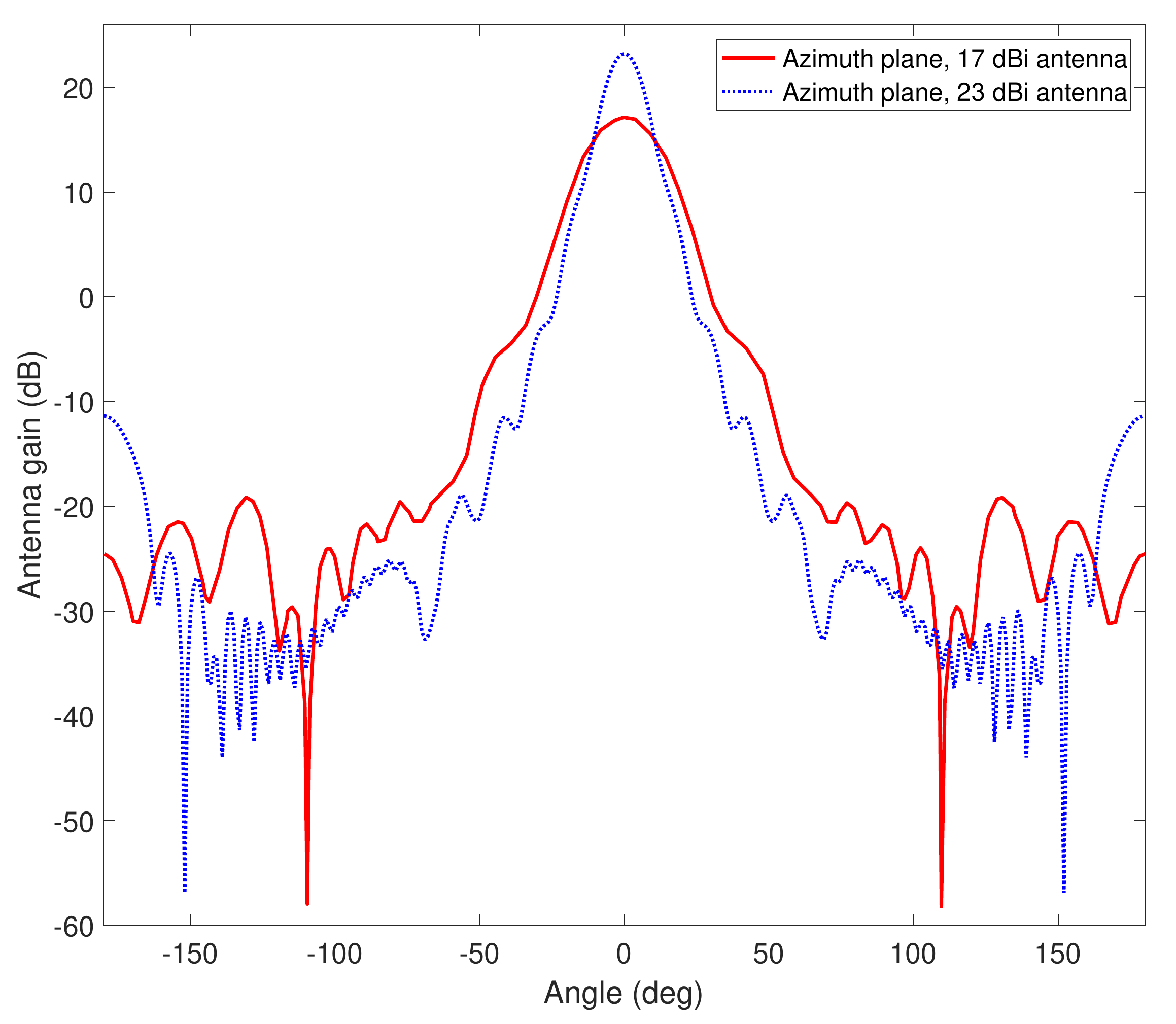}
	  \caption{}  
    \end{subfigure}    
    \begin{subfigure}{\columnwidth}
    \centering
	\includegraphics[width=\textwidth]{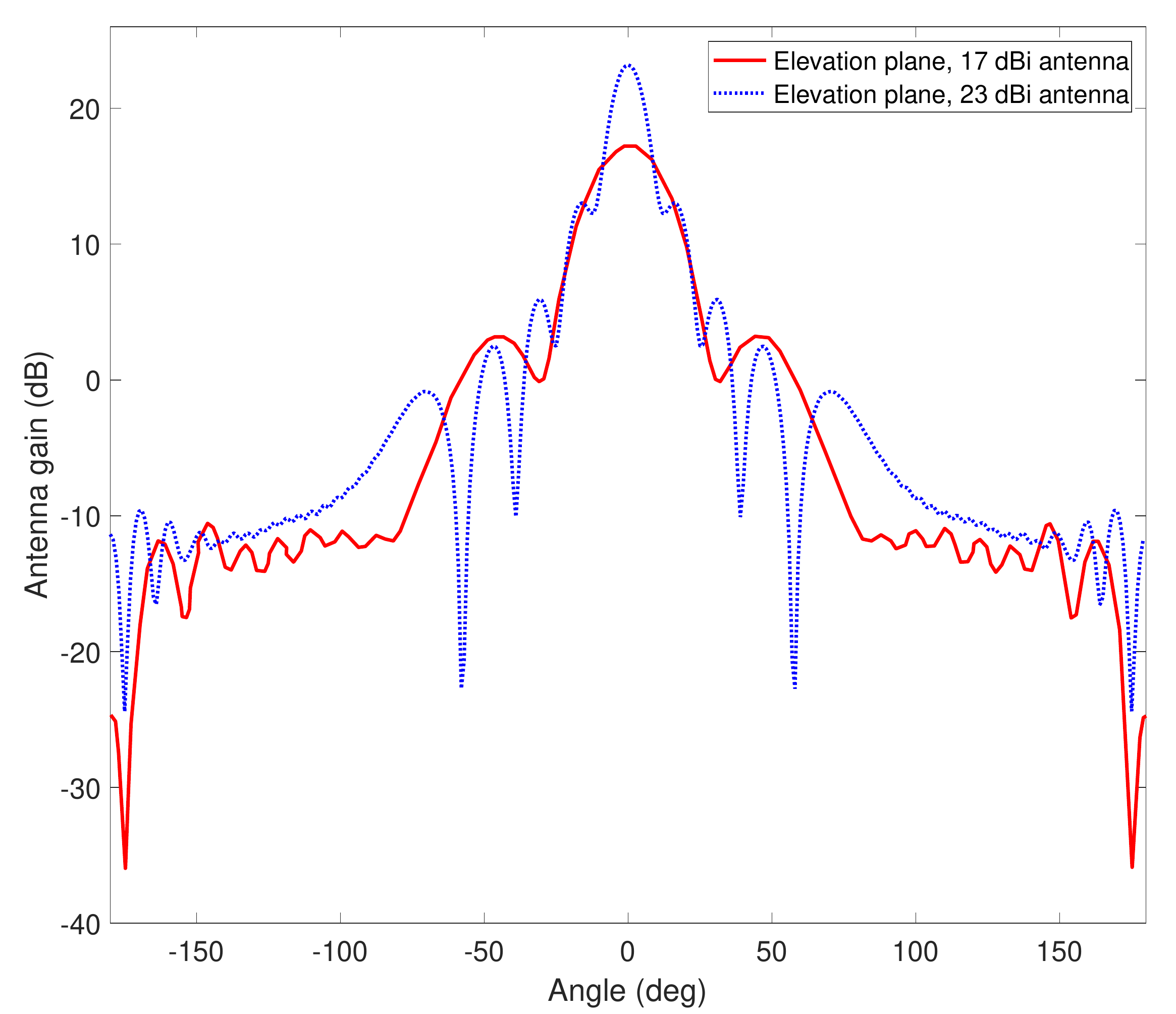}
	  \caption{}  
    \end{subfigure}
    \caption{The antenna radiation pattern for $17$ dBi and $23$ dBi horn antennas in the (a) azimuth plane, (b) elevation plane.}\vspace{-2mm} \label{Fig:Antenna_pattern_17_24}
\end{figure}

\subsection{Ray Tracing and Analytical Simulation Setup}
Ray tracing simulations were carried out using Wireless InSite\textsuperscript{\textregistered} software. The environment model is shown in Fig.~\ref{Fig:ray_tracing_scenario}. The indoor corridor and the outdoor open area were modeled similar to the real environment with as many details as we could. The relative permittivity~(\ref{Eq:Permittivity}) of the concrete floor at $28$~GHz is $5.31$, while it is $3$ for the Armstrong ceiling and $2.94$ for the drywalls. The radiation patterns for the horn antennas shown in Fig.~\ref{Fig:Antenna_pattern_17_24} were obtained from the antenna producer's datasheet.

The analytical simulations for five ray and two rays models from Section~\ref{Section:RX_pwr_model} were conducted based on the geometry of the measurement setup. The reflection coefficient of the materials was obtained from \cite{ITU} at $28$~GHz, the same as used in the ray tracing simulations. Similarly, the antenna gain of the rays at different azimuth and elevation angles were obtained from the antenna data sheets shown in Fig.~\ref{Fig:Antenna_pattern_17_24}. 

\section{Number of Rays and Power of Dominant Rays Obtained Empirically} \label{Section:Num_rays_pow}
\begin{figure}[!t] 
    \centering
	\begin{subfigure}{\columnwidth}
    \centering
	\includegraphics[width=\textwidth]{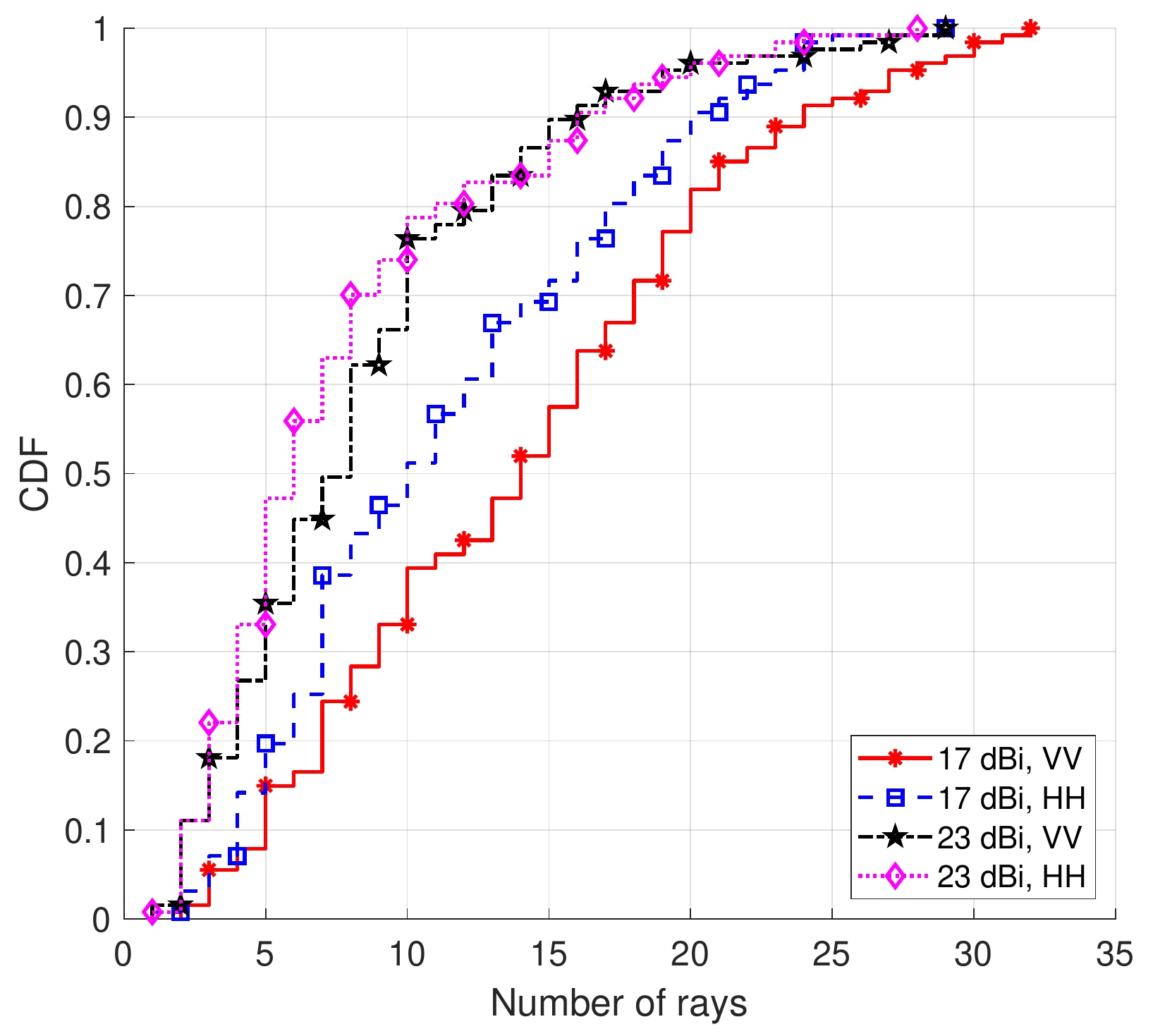}
	  \caption{}  
    \end{subfigure}    
    \begin{subfigure}{\columnwidth}
    \centering
	\includegraphics[width=\textwidth]{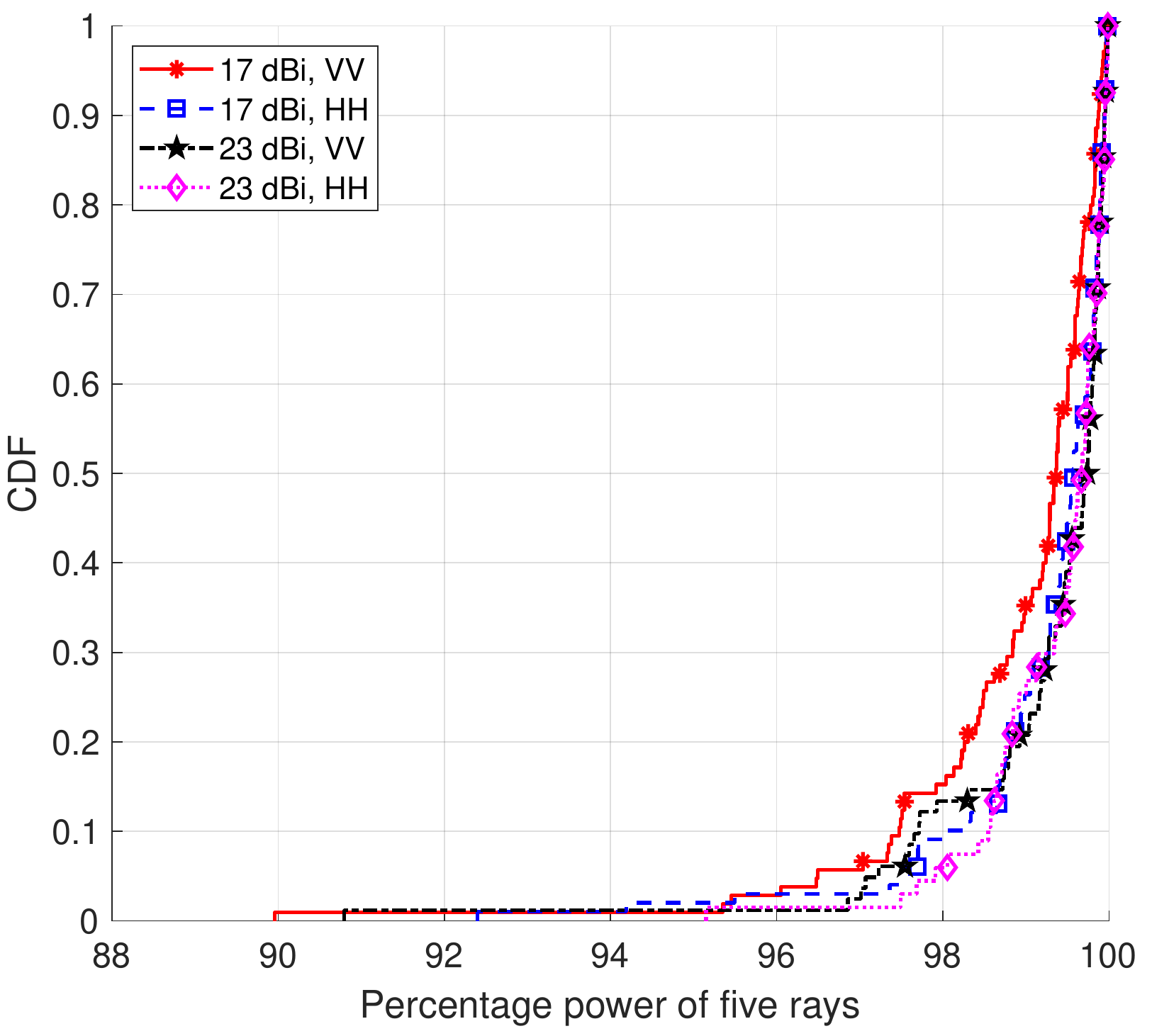}
	  \caption{}  
    \end{subfigure}
   \caption{(a) Number of rays, (b) power of dominant five rays compared to total power presented as percentage, obtained empirically for two antennas and respective antenna orientations.}\vspace{-2mm} \label{Fig:Num_rays_percent_pow}
\end{figure}

The number of rays and the power of the dominant rays are important in deciding the number of rays required for modeling. Fig.~\ref{Fig:Num_rays_percent_pow} shows the empirical cumulative distribution function~(CDF) of the number of rays and power ratio of the sum of the dominant five rays with the total power expressed as percentage for two antennas and their respective orientations.

From Fig.~\ref{Fig:Num_rays_percent_pow}(a), we observe larger number of MPCs for $17$~dBi compared to $23$~dBi. This is due to larger spread of the radiation pattern for $17$~dBi compared to $23$~dBi. Moreover, we observe larger number of rays for VV antenna orientation compared to HH antenna orientation for the two antennas. The difference in the number of rays for VV and HH antenna orientations is larger for $17$~dBi compared to $23$~dBi. This is due to interaction of the antenna radiation in the azimuth and elevation planes with the scatterers and shaped curves~(e.g. sides of the doors) in the environment. This interaction is large for $17$~dBi compared to $23$~dBi due to larger angular spread of the radiation pattern.

Fig.~\ref{Fig:Num_rays_percent_pow}(b) shows the CDF of the empirical power ratio of the sum of the dominant five rays with the total power expressed as a percentage over the link distance. It can be observed that the two antennas with their respective orientations have a percentage of above 90\% over the link distance. For scenarios less than five rays, we have smaller percentages at certain link distances. This proves our claim that five dominant rays are enough to model the received power indoors.

\section{Analysis of Received Power and Path Loss Results} \label{Section:RX_power_PL_analysis}
In this section, analysis and comparison of the empirical received power and path loss results with analytical modeling and results from ray tracing simulations are presented. 

\subsection{Analysis of Received Power Results for 17~dBi Antenna}
\begin{figure}[!t] 
    \centering
	\begin{subfigure}{\columnwidth}
    \centering
	\includegraphics[width=\textwidth]{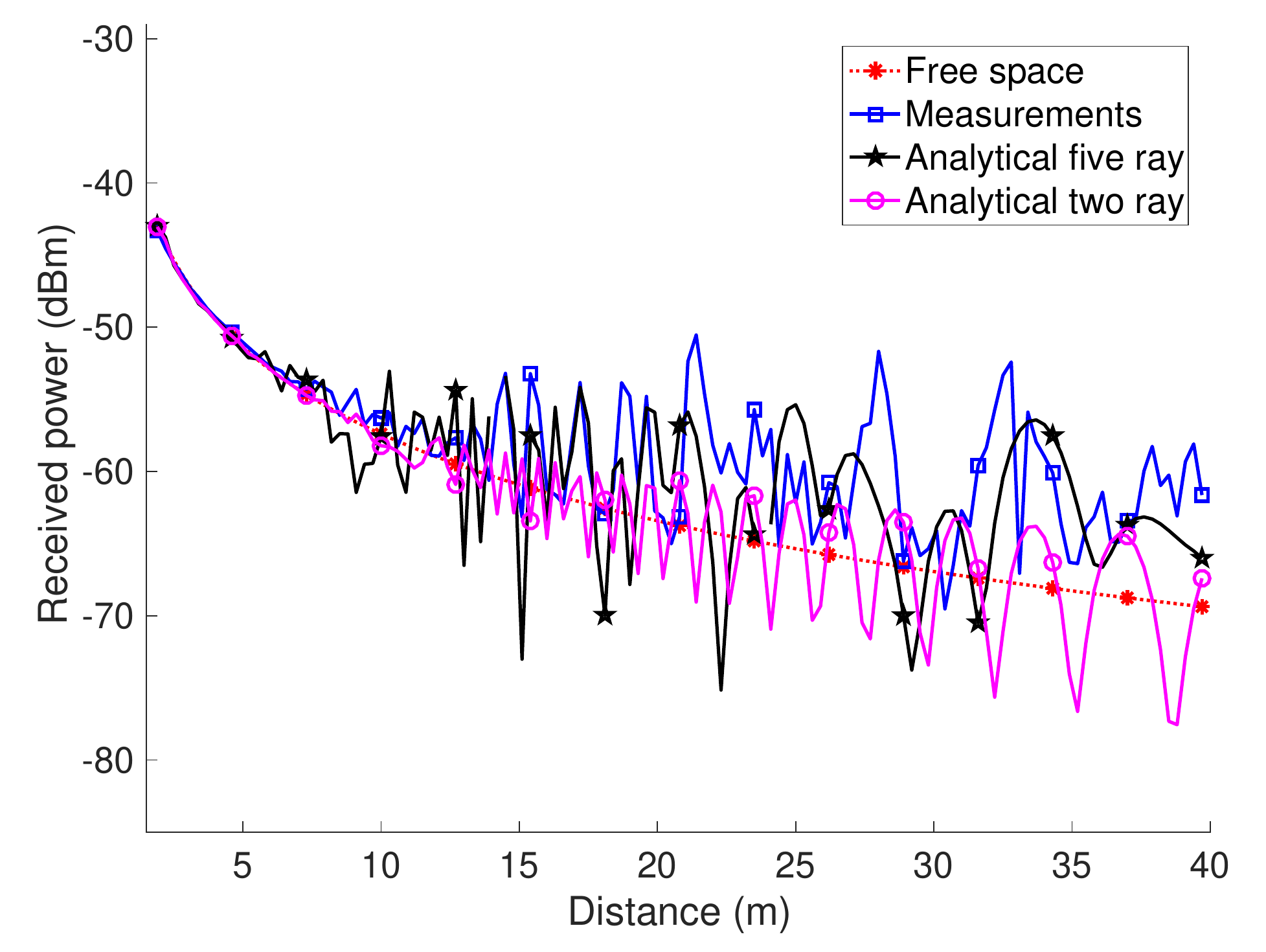}
	  \caption{}  
    \end{subfigure}    
    \begin{subfigure}{\columnwidth}
    \centering
	\includegraphics[width=\textwidth]{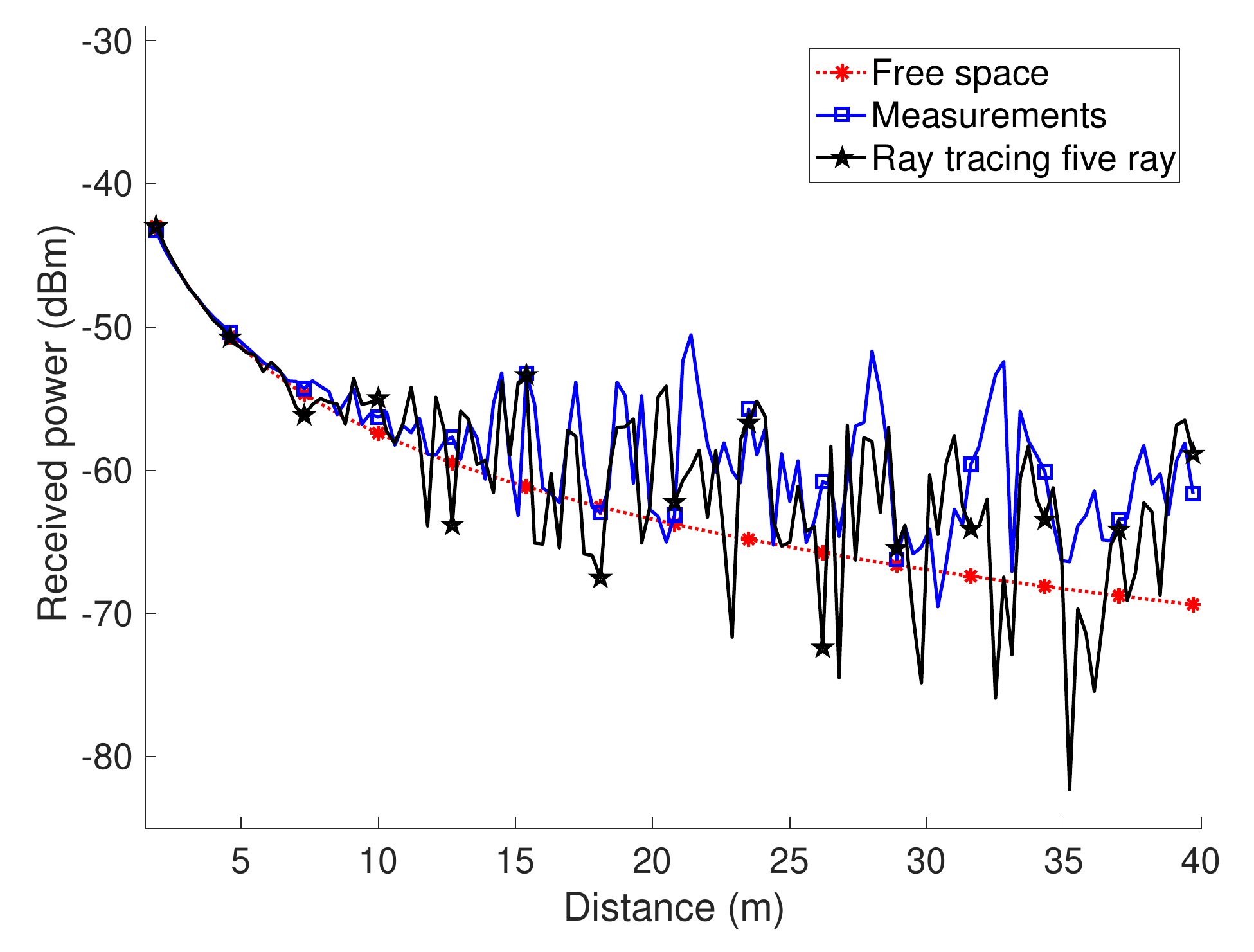}
	  \caption{}  
    \end{subfigure}
    \caption{Received power results for $17$~dBi antenna and VV antenna orientation for indoor corridor plotted against link distance for (a) free space, measurement, five ray and two ray analytical model, (b) free space, measurement and ray tracing simulations.}\vspace{-2mm} \label{Fig:17dBi_VV}
\end{figure}
\begin{figure}[!t] 
    \centering
	\begin{subfigure}{\columnwidth}
    \centering
	\includegraphics[width=\textwidth]{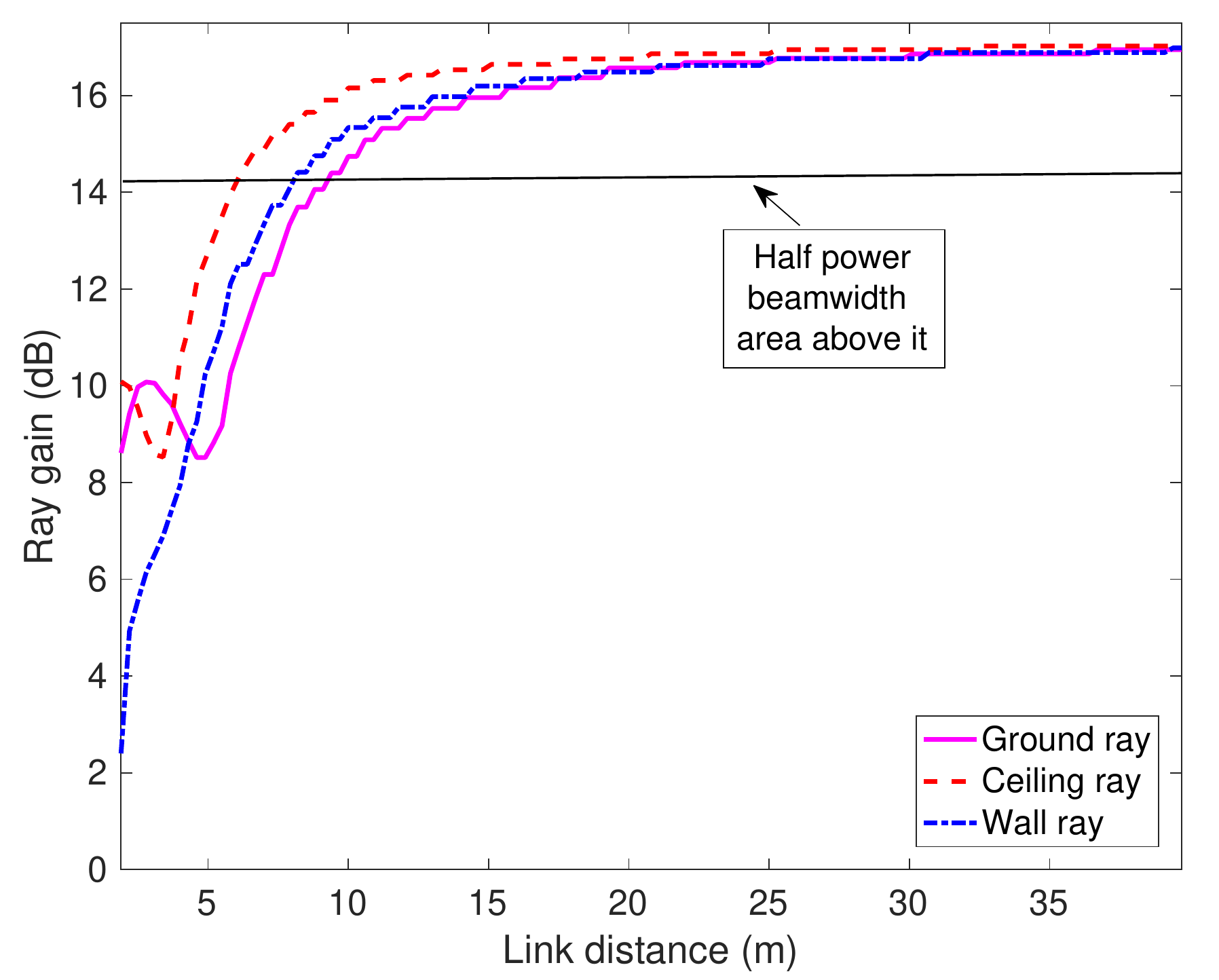}
	  \caption{}  
    \end{subfigure}    
    \begin{subfigure}{\columnwidth}
    \centering
	\includegraphics[width=\textwidth]{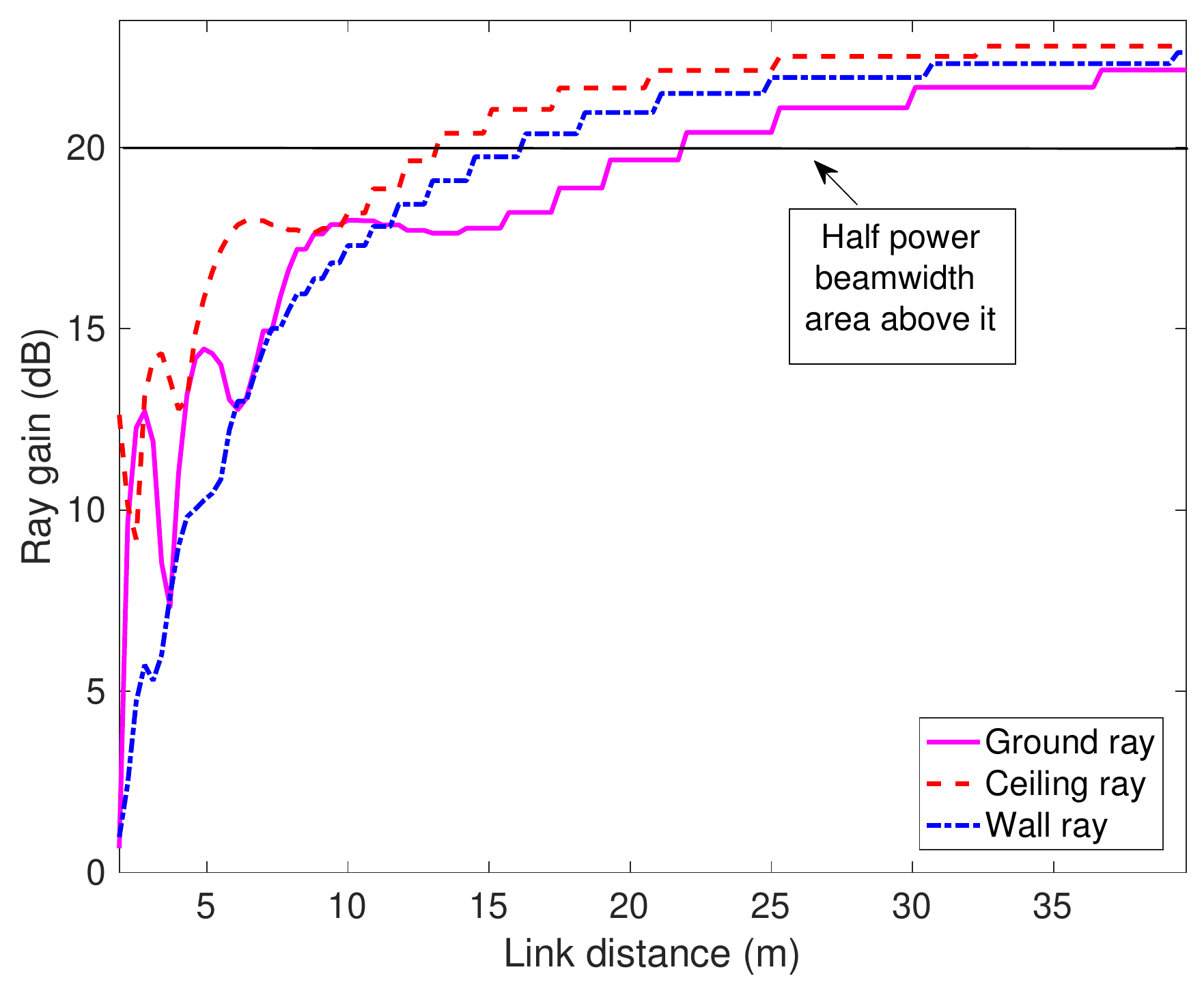}
	  \caption{}  
    \end{subfigure}
    \caption{The gain of rays plotted against the link distance and VV antenna orientation for (a) $17$~dBi horn antenna, (b) $23$~dBi horn antenna.}\vspace{-2mm} \label{Fig:Ray_gains}
\end{figure}

The received power measurement and analytical modeling results for five ray and two ray models for $17$~dBi gain antenna set and VV antenna orientation are shown in Fig.~\ref{Fig:17dBi_VV}(a). The single ray~(free space) received power results are also provided for comparison. It is observed in Fig.~\ref{Fig:17dBi_VV}(a) that the received power behaves the same as the free space attenuation for the first $6$~m. This is because the reception of the reflected rays is small due to the large $\Psi$ angle. Therefore, their contribution to the total received power is small. However, as the link distance increases, the reception of the reflected rays increases and their contribution to the overall received power also increases shown in Fig.~\ref{Fig:Ray_gains}(a). This results in peaks and dips of the received power at different link distances due to the coherent addition of the rays (see Section~\ref{Section:Rec_pwr_five_ray}). 

Comparing the received power results of five ray and two ray analytical models with the measurements in Fig.~\ref{Fig:17dBi_VV}(a), it can be observed that five ray model closely follows the measurement results compared to two ray. For the two ray model, the mean is close to the free space. However, the contribution of additional rays causes higher received power peaks and valleys for five ray. A ray model above five is not introduced here, because it will introduce higher complexity and more dependence on the surrounding environment. A fitting comparison of the five ray and two ray models with the measurements is provided in Section~\ref{Section:Five_two_compare}. 

\begin{figure}[!t] 
    \centering
	\begin{subfigure}{\columnwidth}
    \centering
	\includegraphics[width=\textwidth]{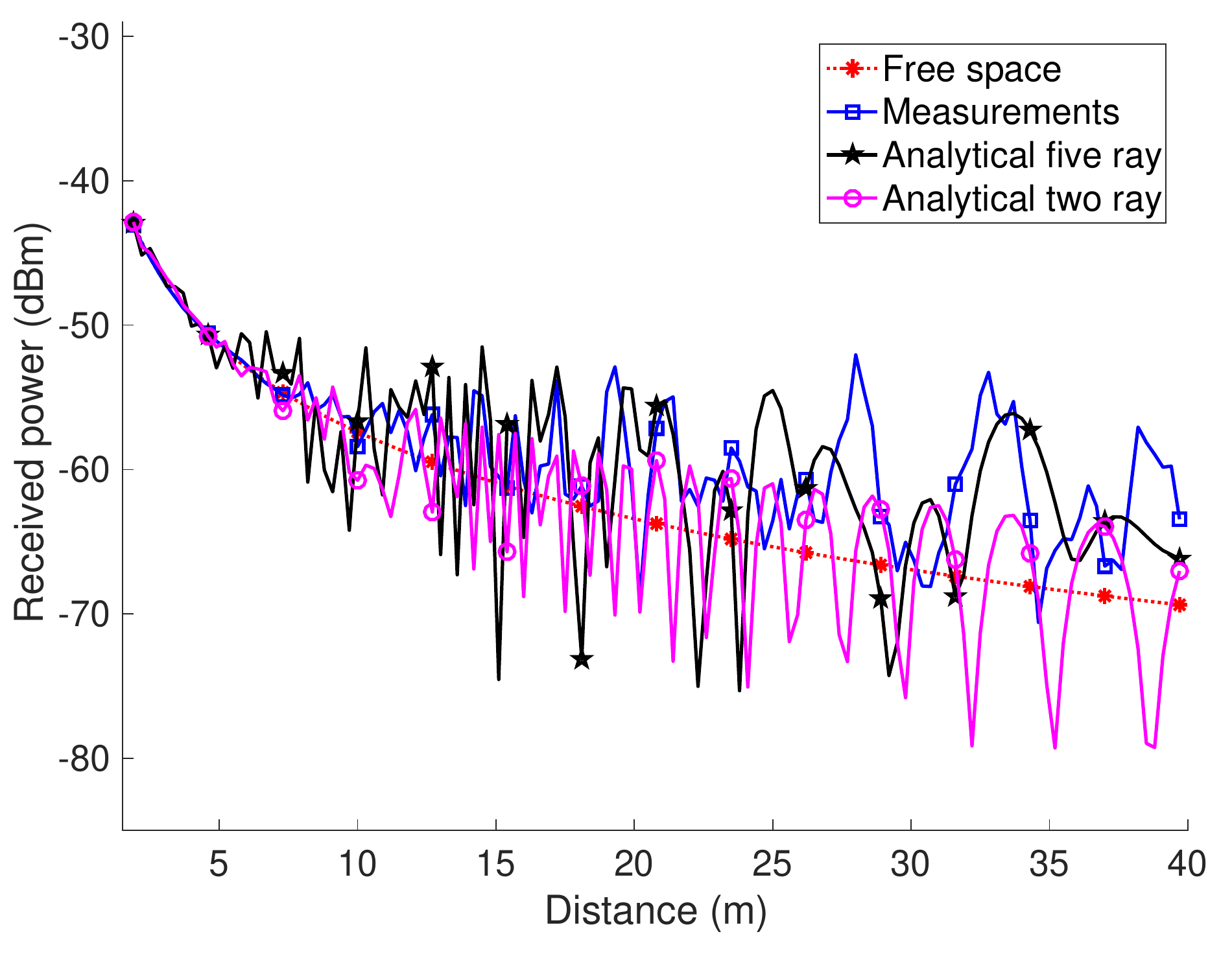}
	  \caption{}  
    \end{subfigure}    
    \begin{subfigure}{\columnwidth}
    \centering
	\includegraphics[width=\textwidth]{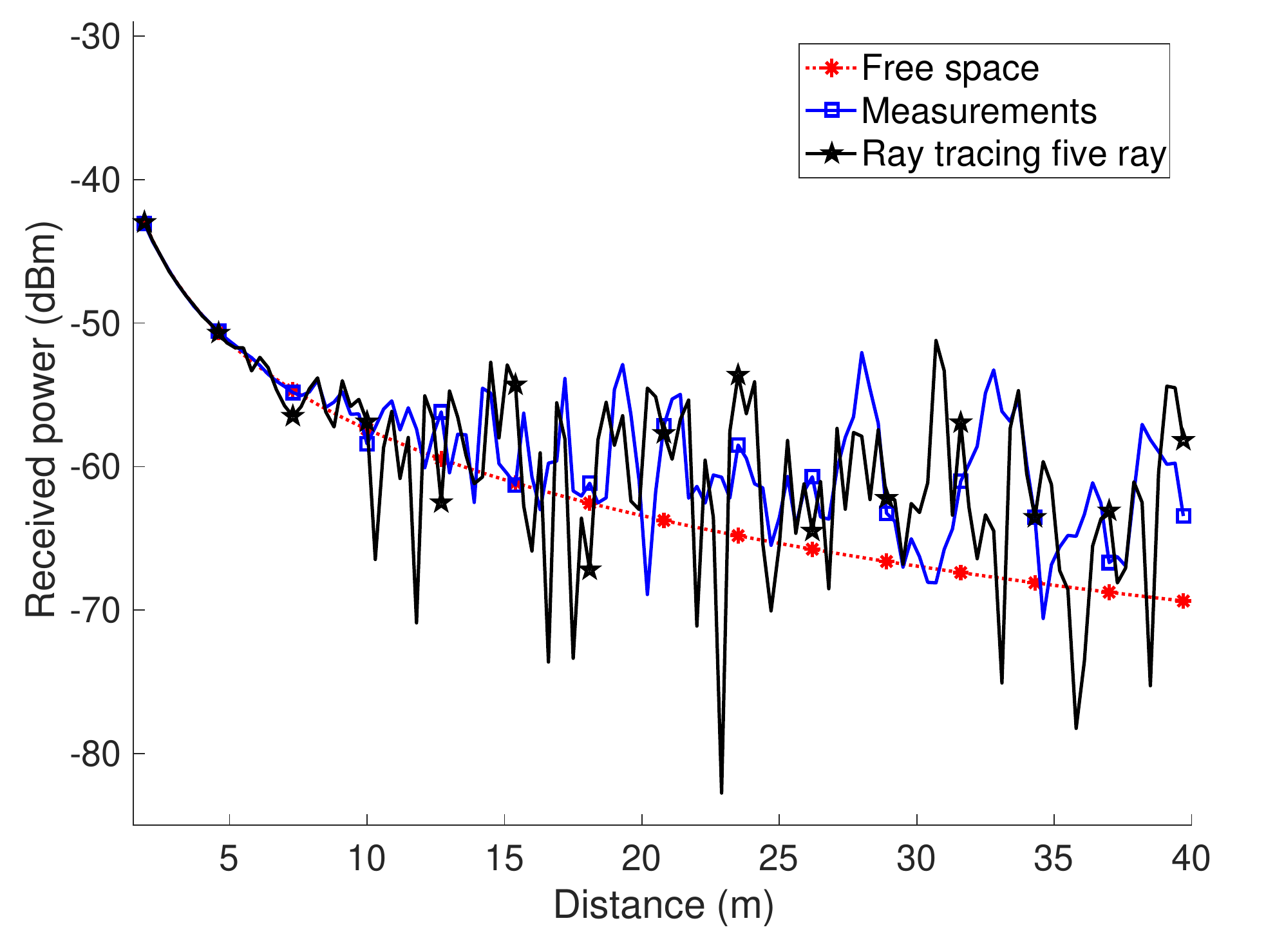}
	  \caption{}  
    \end{subfigure}
    \caption{Received power results for $17$~dBi antenna and HH antenna orientation for indoor corridor plotted against link distance for (a) free space, measurement, five ray and two ray analytical model, (b) free space, measurement and ray tracing simulations.}\vspace{-2mm} \label{Fig:17dBi_HH}
\end{figure}

In overall, the difference between the measurement and analytical results are mainly due to 1) taking only first-order reflections into account, 2) additional scatterers and reflectors in the real environment i.e. measuring equipment, metallic doors, and ceiling metal holders, which are not considered in analytical modeling, and 3) cross polarization of reflected components in the real-world (assumed to be negligible~($|\pmb{\rho}^{\rm (TX)}_{i}\cdot\pmb{\rho}^{\rm (RX)}_{i}|=1$) in analytical models).  

The measurement and Wireless InSite\textsuperscript{\textregistered} ray tracing simulation results for five ray model are shown in Fig.~\ref{Fig:17dBi_VV}(b). Simulation results using two ray model gave quite similar results as the analytical two ray calculations, so we preferred to omit for this figure. Similar to the analytical five ray model, simulation results mean value show a close fit to the measurements. The mismatch of peaks and dips are due to similar reasons stated for analytical calculations above. Even though, the ray tracing environment has been created in the software as detailed as possible by introducing metallic doors and other physical shapes in the corridor, the properties of materials, corners, and edges of the structures and diffuse scattering of real-world objects cannot be exactly imitated.

		
            

\begin{figure}[!t] 
    \centering
	\begin{subfigure}{\columnwidth}
    \centering
	\includegraphics[width=\textwidth]{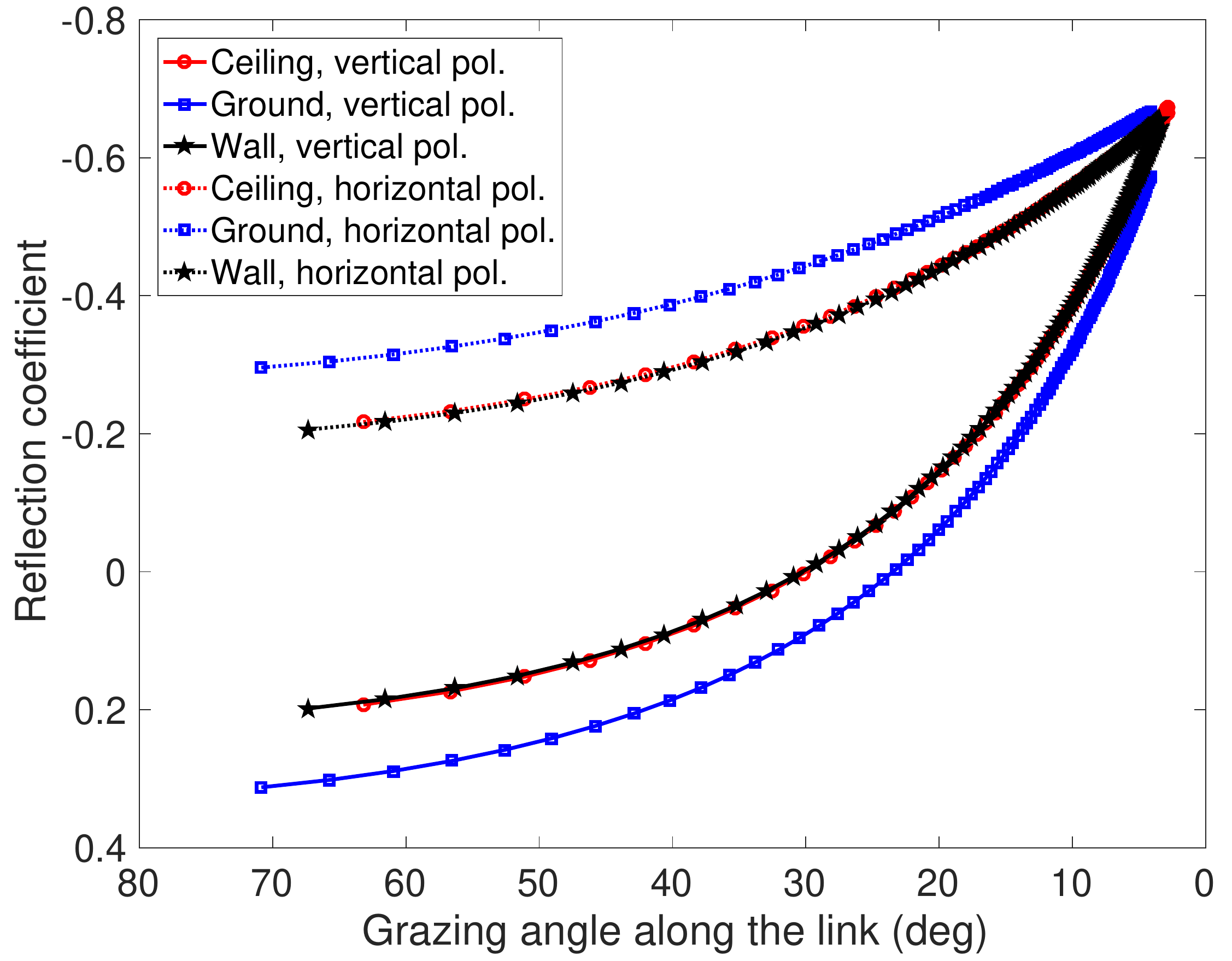}
	  \caption{}  
    \end{subfigure}    
    \begin{subfigure}{\columnwidth}
    \centering
	\includegraphics[width=\textwidth]{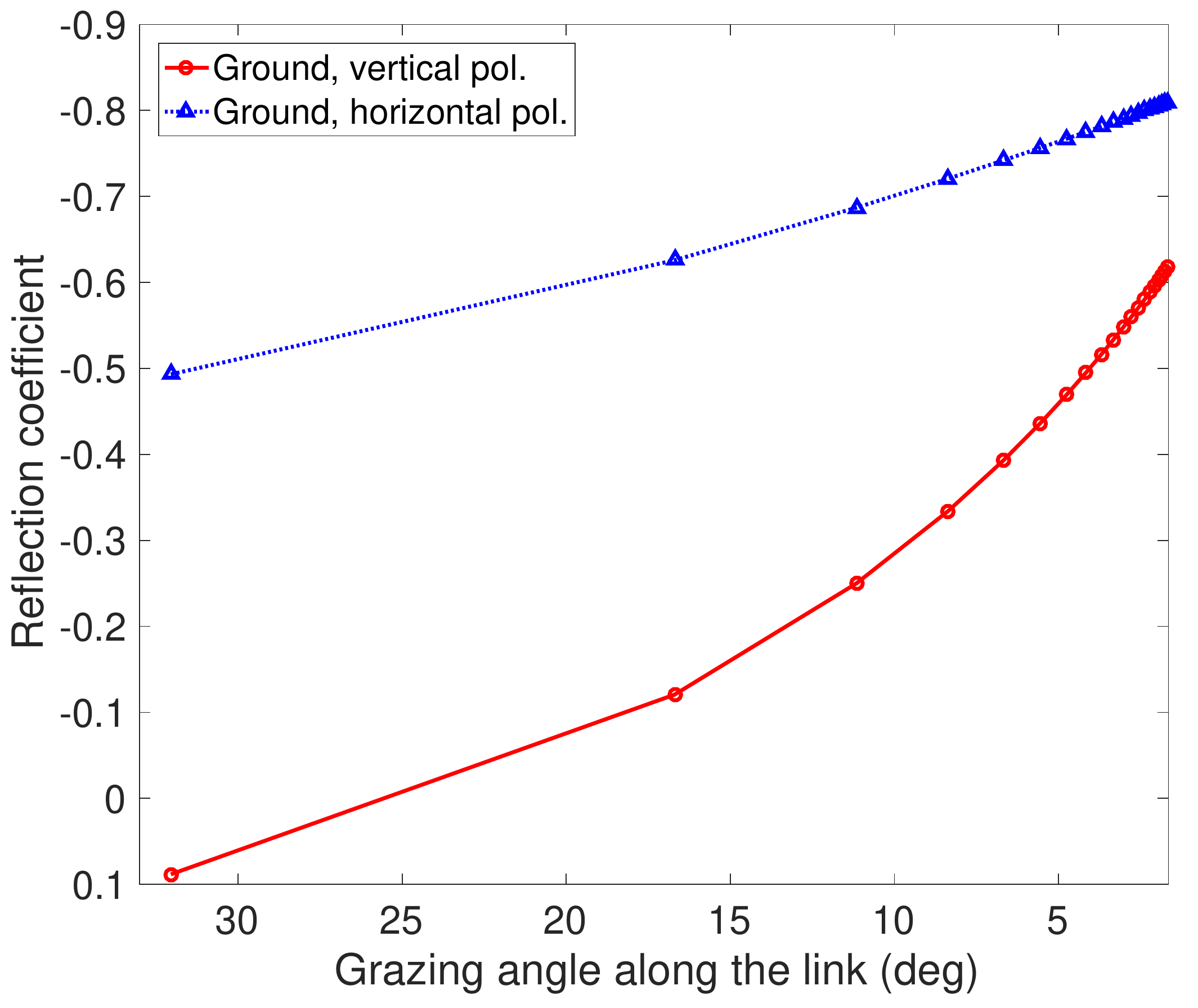}
	  \caption{}  
    \end{subfigure}
    \caption{The Fresnel reflection coefficient of different materials for vertical and horizontal polarization plotted against the grazing angle for (a) indoor corridor scenario, (b) outdoor open area scenario.}\vspace{-4mm} \label{Fig:Reflection_coeff}
\end{figure}

The received power results from measurements, five ray and two ray analytical calculations for $17$~dBi antenna set at HH antenna orientation are shown in Fig.~\ref{Fig:17dBi_HH}(a). Similarly, the measurement and ray tracing simulation results are shown in Fig.~\ref{Fig:17dBi_HH}(b). Comparing Fig.~\ref{Fig:17dBi_VV} and Fig.~\ref{Fig:17dBi_HH}, it can be observed that there are no significant changes between VV and HH schemes. This is due to symmetry in the antenna patterns at the azimuth and elevation planes. The difference observed only comes from Fresnel reflection coefficients' dependence on wave polarization shown in Fig.~\ref{Fig:Reflection_coeff}(a).

\begin{figure}[!t] 
    \centering
	\begin{subfigure}{\columnwidth}
    \centering
	\includegraphics[width=\textwidth]{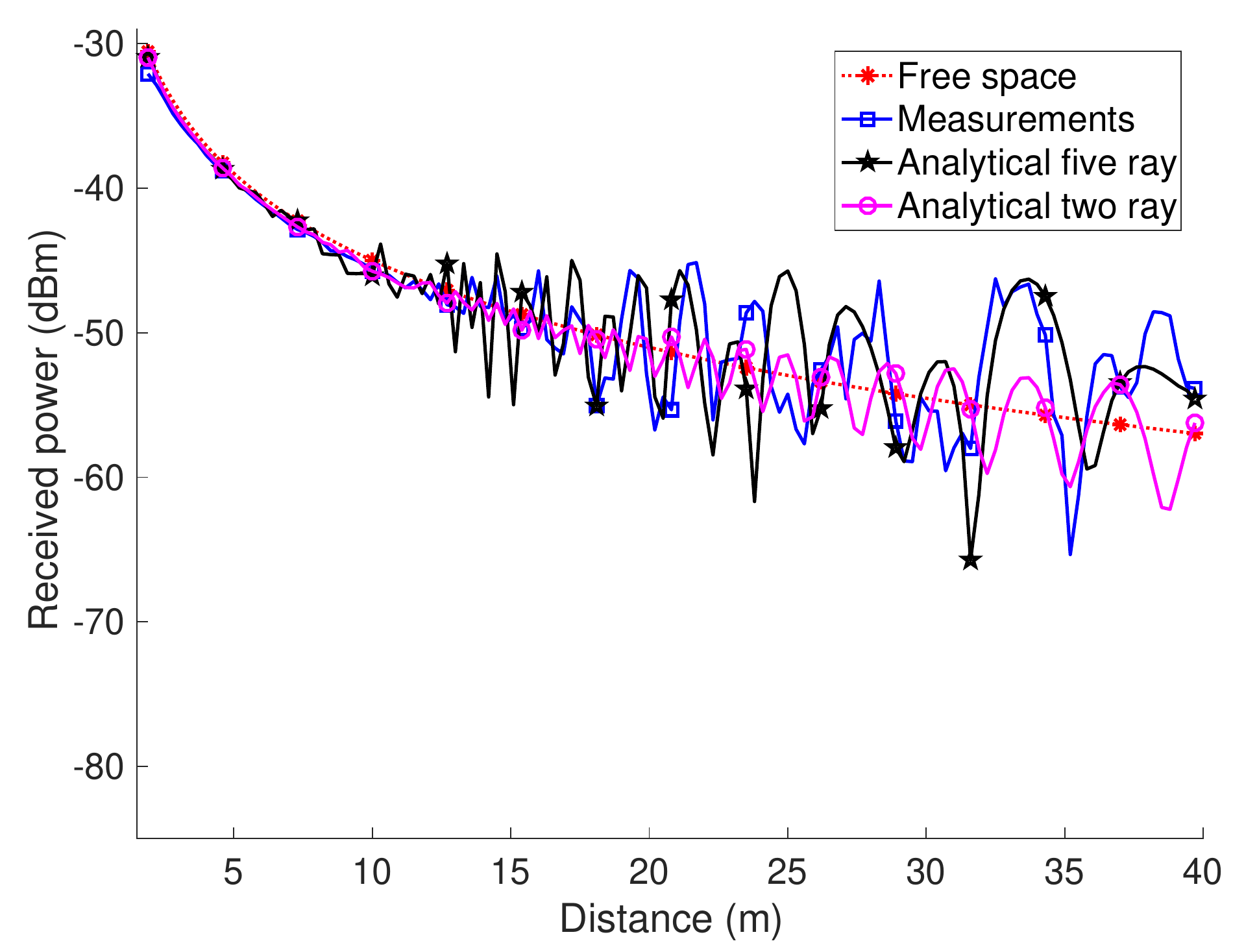}
	  \caption{}  
    \end{subfigure}    
    \begin{subfigure}{\columnwidth}
    \centering
	\includegraphics[width=\textwidth]{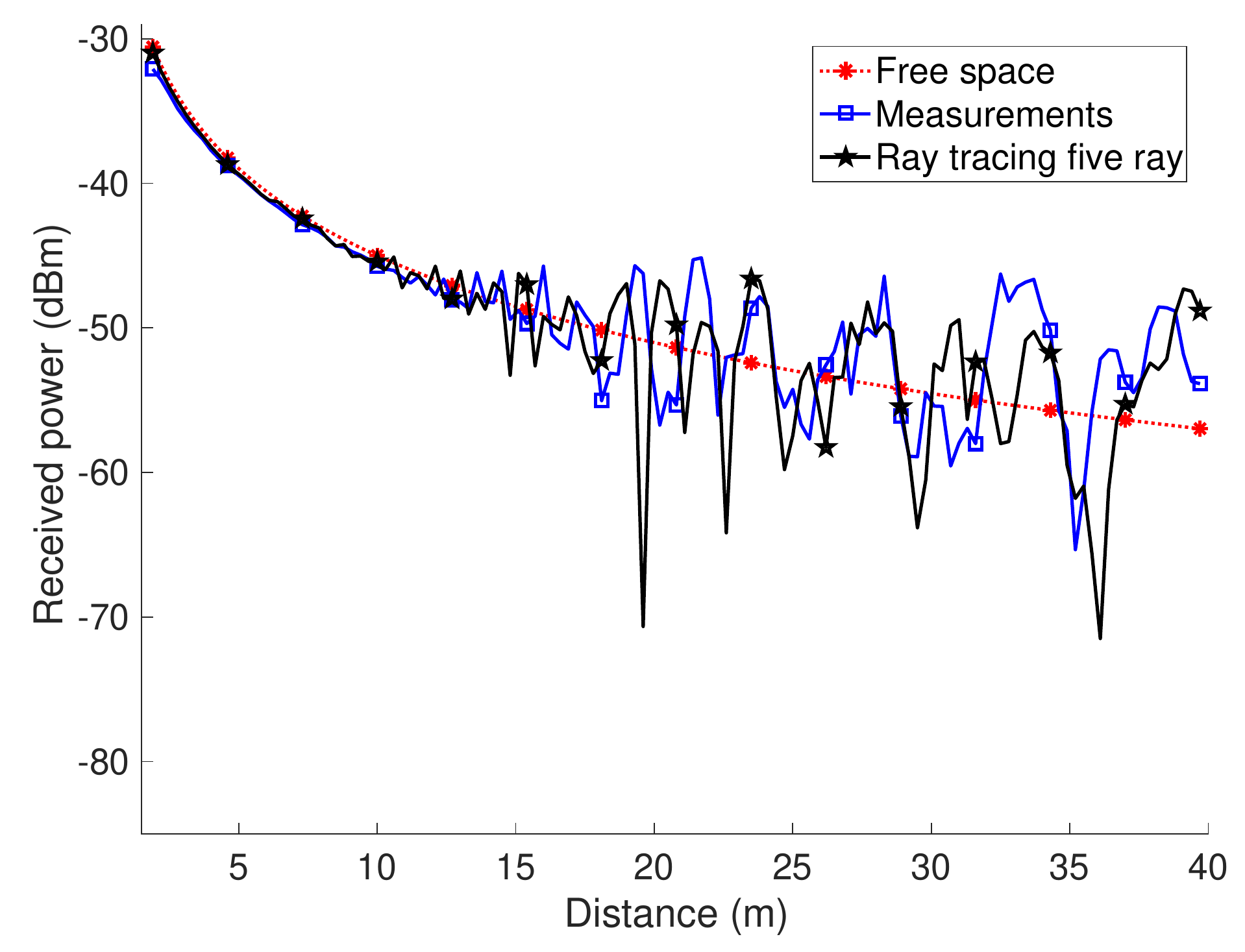}
	  \caption{}  
    \end{subfigure}
    \caption{Received power results for $23$~dBi antenna and VV antenna orientation for indoor corridor plotted against the link distance for (a) free space, measurement, five ray and two ray analytical model, (b) free space, measurement and ray tracing simulations.}\vspace{-2mm} \label{fig:24dBi_VV}
\end{figure}
\begin{figure}[!t] 
    \centering
	\begin{subfigure}{\columnwidth}
    \centering
	\includegraphics[width=\textwidth]{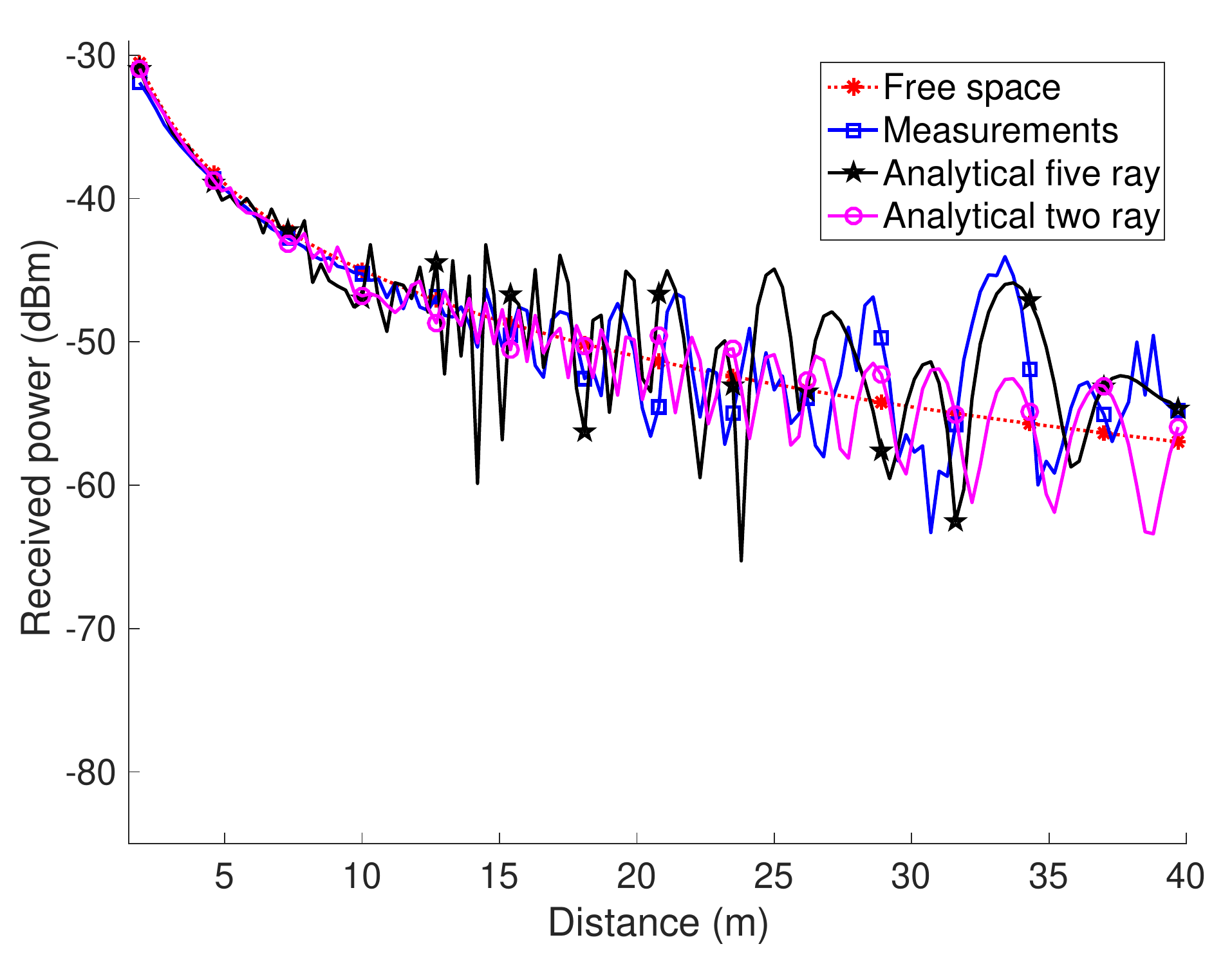}
	  \caption{}  
    \end{subfigure}    
    \begin{subfigure}{\columnwidth}
    \centering
	\includegraphics[width=\textwidth]{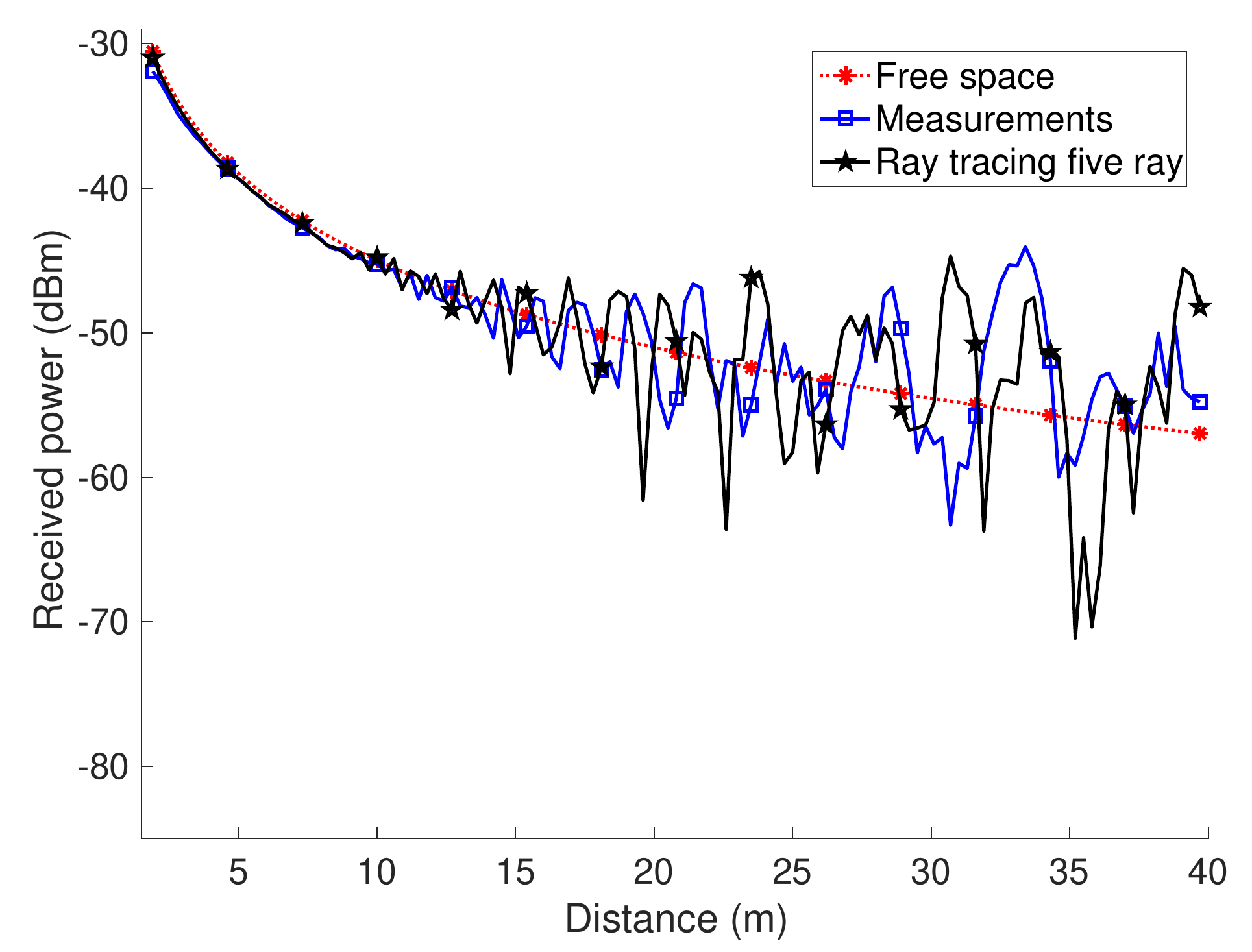}
	  \caption{}  
    \end{subfigure}
    \caption{Received power results for $23$~dBi antenna and HH antenna orientation for indoor corridor plotted against the link distance for (a) free space, measurement, five ray and two ray analytical model, (b) free space, measurement and ray tracing simulations.}\vspace{-2mm} \label{Fig:24dBi_HH}
\end{figure}

\subsection{Analysis of Received Power Results for 23~dBi Antenna}
The received power measurement and analytical modeling results for five ray and two ray models for $23$~dBi gain antenna set and VV antenna orientation are shown in Fig.~\ref{fig:24dBi_VV}(a). It is observed in Fig.~\ref{Fig:17dBi_VV}(a) that the received power behaves the same as the free space attenuation for the first $10$~m. This distance is larger compared to as observed for the $17$~dBi antennas. This is mainly due to the small antenna gain of reflected rays for $23$~dBi at small link distance shown \vspace{-5mm}in Fig.~\ref{Fig:Ray_gains}. \vspace{4mm} The reflected rays have to travel larger link distance compared to the $17$~dBi antenna to approach the half-power antenna gain region $A_{\rm hpa}$ (from Section~\ref{Section:Antenna_rad_pattern}). Moreover, we have larger fluctuations of the antenna gain at smaller link distances compared to $17$~dBi antenna because of the antenna pattern in the elevation plane (Fig.~\ref{Fig:Ray_gains}).

At Fig.~\ref{fig:24dBi_VV}(a), it can be observed that the analytical five ray model provides a closer match to $23$~dBi antenna measurement results compared to $17$~dBi. This can be attributed to the limited effects of the surroundings on the propagation due to small beamwidth. Similar to $17$~dBi antenna, the two ray model does not provide close-fitting to the measurement results. The ray tracing simulation results for five ray model and $23$~dBi antenna with VV antenna orientation measurements are shown in Fig.~\ref{fig:24dBi_VV}(b). The ray tracing and measurement results are also close to each other. The measurement, analytical five ray and two ray and ray tracing simulation results for HH antenna orientation are provided in Fig.~\ref{Fig:24dBi_HH}. We get similar results for two different antenna orientation schemes as expected.  

\subsection{Analysis of Received Power Results for Outdoor Open Area}
\begin{figure}[!t]
\centering
\centerline{\includegraphics[width=\columnwidth]{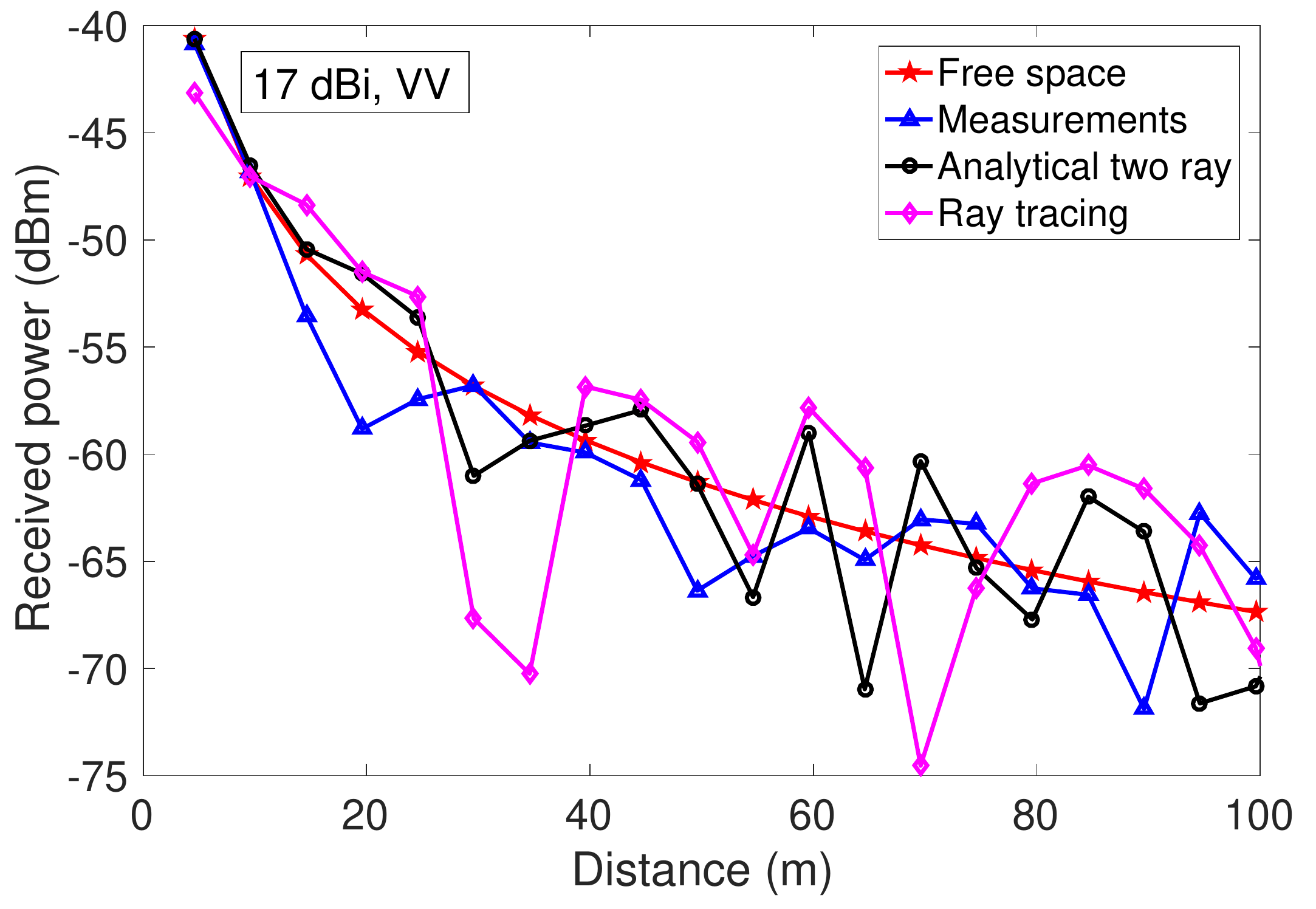}}
\caption{Received power for open area outdoor plotted against the link distance for free space, measurement, analytical and ray tracing simulation results based on two ray model for VV antenna orientation. } \label{Fig:Rec_pwr_two_ray_outdoor}
\end{figure}

The received power results for the outdoor open area are shown in Fig.~\ref{Fig:Rec_pwr_two_ray_outdoor} for $17$~dBi antenna and VV orientation. The results for the $23$~dBi antenna show similar trend as for $17$~dBi. However, similar to indoor results, we have closer match with the two ray model for the $23$~dBi antenna compared to $17$~dBi antenna. The Fresnel reflection coefficient used for the analytical results is shown in Fig.~\ref{Fig:Reflection_coeff}(b). It can be observed that the measured received power fluctuates around the free space. Besides, the measured power has limited contribution from the reflected rays and mainly the GRC is the dominant reflected ray as the analytical and ray tracing simulation results also follow the measurements. 

\begin{table}[!t]
\caption{XPD Factor for $17$~dBi and $23$~dBi Antenna Gains and for VV and HH Antenna Orientations. \label{Table:polarization loss}}
{\begin{tabular*}{20pc}{@{\extracolsep{\fill}}lcccc}\toprule
Parameters&17 dBi, VV & 17 dBi, HH  & 23 dBi, VV& 23 dBi, HH\\
\midrule
$\rho$ &22.6 &22.1 & 29.3 & 29.1 \\
\bottomrule
\end{tabular*}}{} 
\end{table}

\subsection{Cross Polarization Discrimination Results}
The XPD factor~(see Section~\ref{Section:XPD}) results are shown in Table~\ref{Table:polarization loss}. The majority of the cross polarization components for the VH antenna orientation arise due to reflections from the surroundings. Since $17$~dBi antennas receive reflected rays better in comparison to $23$~dBi antennas in the corridor environment, we observe a smaller XPD factor for $17$~dBi than $23$~dBi gain antenna. Moreover, there is no significant change in the XPD factor from VV to HH for both antenna sets. 

\section{Path Loss Modeling Results and Comparison of Five Ray and Two Ray Models}
\label{Section:Five_two_compare}
In this section, path loss model results~(discussed in Section~\ref{Section:Path_loss}) obtained through measurements, five ray and two ray analytical models and ray tracing simulations are discussed. Moreover, a comparison of the five ray and two ray models is provided based on the path loss model parameters. 
\begin{table*}[!h]
\renewcommand{\arraystretch}{1.3}
\caption{Path loss fitting parameters for indoor corridor and different antenna gains and respective orientations for measurement, analytical~(five and two ray) and ray tracing simulations~(five ray).}
\centering
\resizebox{\textwidth}{!}{
\begin{tabular}{|p{.07cm}|p{0.07cm}|p{0.07cm}|p{0.07cm}|p{0.07cm}|p{0.07cm}|p{0.07cm}|p{0.07cm}|p{0.07cm}|p{0.07cm}|p{0.07cm}|p{0.07cm}|p{0.07cm}|p{0.07cm}|p{0.07cm}|p{0.07cm}|p{0.07cm}|} 
\toprule
		\multicolumn{1}{p{.3cm}}{}&\multicolumn{4}{p{3.5cm}}{{~~~~~~~~~~~~~~~~~~~~~17~dBi, VV}}&\multicolumn{4}{p{3.5cm}}{{~~~~~~~~~~~~~~~~~~~17~dBi, HH}}&\multicolumn{4}{p{3.5cm}}{{~~~~~~~~~~~~~~~~~~~23~dBi, VV}}&\multicolumn{4}{p{3.5cm}}{{~~~~~~~~~~~~~~~~~~~23~dBi, HH}} \\
		
\hline
             \multicolumn{1}{p{.8cm}}{{Param.}}&\multicolumn{1}{p{.7cm}}{{Meas.}}&\multicolumn{1}{p{.9cm}}{{Analyt. five ray}}&\multicolumn{1}{p{.9cm}}{{Analyt. two ray}}&\multicolumn{1}{p{1.2cm}}{{RT. sim. five ray}}&\multicolumn{1}{p{.7cm}}{{Meas.}}&\multicolumn{1}{p{.9cm}}{{Analyt. five ray}}&\multicolumn{1}{p{.9cm}}{{Analyt. two ray}}&\multicolumn{1}{p{1.2cm}}{{RT. sim. five ray}}&\multicolumn{1}{p{.7cm}}{{Meas.}}&\multicolumn{1}{p{.9cm}}{{Analyt. five ray}}&\multicolumn{1}{p{.9cm}}{{Analyt. two ray}}&\multicolumn{1}{p{1.2cm}}{{RT. sim. five ray}}&\multicolumn{1}{p{.7cm}}{{Meas.}}&\multicolumn{1}{p{.9cm}}{{Analyt. five ray}} &\multicolumn{1}{p{.9cm}}{{Analyt. two ray}}&\multicolumn{1}{p{1.2cm}}{{RT. sim. five ray}}\\
\hline

\multicolumn{1}{c}{$\alpha$}&\multicolumn{1}{c}{{-1.26}}&\multicolumn{1}{c}{{-1.46}}&\multicolumn{1}{c}{{-2.07}}&\multicolumn{1}{c}{{-1.64}}&\multicolumn{1}{c}{{-1.32}}&\multicolumn{1}{c}{{-1.45}}&\multicolumn{1}{c}{{-2.01}}&\multicolumn{1}{c}{{-1.4}}&\multicolumn{1}{c}{{-1.64}}&\multicolumn{1}{c}{{-1.7}}&\multicolumn{1}{c}{{-2.02}}&\multicolumn{1}{c}{{-1.86}}&\multicolumn{1}{c}{{-1.71}}&\multicolumn{1}{c}{{-1.62}}&\multicolumn{1}{c}{{-2.01}}&\multicolumn{1}{c}{{-1.8}}\\
\hline

\multicolumn{1}{c}{\thead{$\beta$}}&\multicolumn{1}{c}{{-42.64}}&\multicolumn{1}{c}{{-42}}&\multicolumn{1}{c}{{-37.32}}&\multicolumn{1}{c}{{-40}}&\multicolumn{1}{c}{{-42.4}}&\multicolumn{1}{c}{{-41.4}}&\multicolumn{1}{c}{{-37.5}}&\multicolumn{1}{c}{{-42.3}}&\multicolumn{1}{c}{{-28.7}}&\multicolumn{1}{c}{{-28.2}}&\multicolumn{1}{c}{{-25.2}}&\multicolumn{1}{c}{{-26.7}}&\multicolumn{1}{c}{{-28}}&\multicolumn{1}{c}{{-28.7}}&\multicolumn{1}{c}{{-25.3}}&\multicolumn{1}{c}{{-27}}\\
\hline

\multicolumn{1}{c}{\thead{$\sigma$~(dB)}}&\multicolumn{1}{c}{{3.4}}&\multicolumn{1}{c}{{4.8}}&\multicolumn{1}{c}{{2.89}}&\multicolumn{1}{c}{{4.5}}&\multicolumn{1}{c}{{3.4}}&\multicolumn{1}{c}{{4.7}}&\multicolumn{1}{c}{{4.06}}&\multicolumn{1}{c}{{5.3}}&\multicolumn{1}{c}{{3.3}}&\multicolumn{1}{c}{{3.5}}&\multicolumn{1}{c}{{1.62}}&\multicolumn{1}{c}{{4.1}}&\multicolumn{1}{c}{{3.2}}&\multicolumn{1}{c}{{3.8}}&\multicolumn{1}{c}{{2.15}}&\multicolumn{1}{c}{{4.3}}\\
 \bottomrule
		\end{tabular}
\label{Table:Path_loss}
}
\vspace{2mm}
\end{table*}

\begin{table*}[!t]
\caption{$z$ values obtained from (\ref{Eq:z_value}) for slope of the path loss linear fittings.  \label{Table:z_slope}}
{\begin{tabular*}{\textwidth}{@{\extracolsep{\fill}}lccc}\toprule
Scenario & $z$ value for analytical (two ray) & $z$ value for analytical (five ray)  & $z$ value for ray tracing (five ray) \\
\midrule
17~dBi, VV & 6.07 & 1.23 & 2.42 \\
17~dBi, HH & 4.69 & 0.836 & 0.45 \\
23~dBi, VV & 3.64 & 0.243 & 1.46 \\
23~dBi, HH & 2.82 & -0.614 & 0.643 \\
\bottomrule
\end{tabular*}}
\vspace{2mm}
\end{table*}

\begin{table*}[!t]
\caption{$z$ values obtained for y-intercept of the path loss linear fittings.  \label{Table:z_intercept}}
{\begin{tabular*}{\textwidth}{@{\extracolsep{\fill}}lccc}\toprule
Scenario & $z$ value for analytical (two ray) & $z$ value for analytical (five ray)  & $z$ value for ray tracing (five ray) \\
\midrule
17~dBi, VV & -3.35 & -0.34 & -1.42 \\
17~dBi, HH & -2.61 & -0.5 & -0.6 \\
23~dBi, VV & -2.64 & -0.278 & -1.06 \\
23~dBi, HH & -2.02 & 0.323 & -0.543 \\
\bottomrule
\end{tabular*}}{} 
\end{table*}

\subsection{Path Loss Modeling Results Analysis}
The path loss model parameters (from Section~\ref{Section:Path_loss}) for the indoor corridor are shown in Table~\ref{Table:Path_loss}. The slope of the path loss $\alpha$ is smaller compared to the free space for both $17$~dBi and $23$~dBi gain antennas for measurements and five ray model. However, the slope is similar to the free space for the two ray model. 

It is observed in Table~\ref{Table:Path_loss} that the path loss slope $\alpha$ is smaller for the $17$~dBi antenna than the $23$~dBi antenna. This is because reflected rays are received better by the $17$~dBi antennas due to their larger beamwidth. For the $23$~dBi gain antenna, the slope $\alpha$ is larger than $17$~dBi antenna, however, still smaller than the free space~(single ray). This indicates that additionally reflected rays contribute considerably to the overall received power, even with small antenna beamwidth.

For outdoor open area, the path loss parameters are $\alpha = -1.87,~ -2.11,~ -1.75$, $\beta=-30.5,~ -25.94,~ -31.46$, and $\sigma=2.4,~3.2,~5.2$ for measurements, analytical calculations and ray tracing simulations, respectively. The slopes are slightly less than the slope of the free space attenuation curve. This is because, similar to the indoor corridor environment, the contribution of reflected rays are mainly from the GRC.
\begin{figure} 
    \centering
	\begin{subfigure}{\columnwidth}
    \centering
	\includegraphics[width=0.9\textwidth]{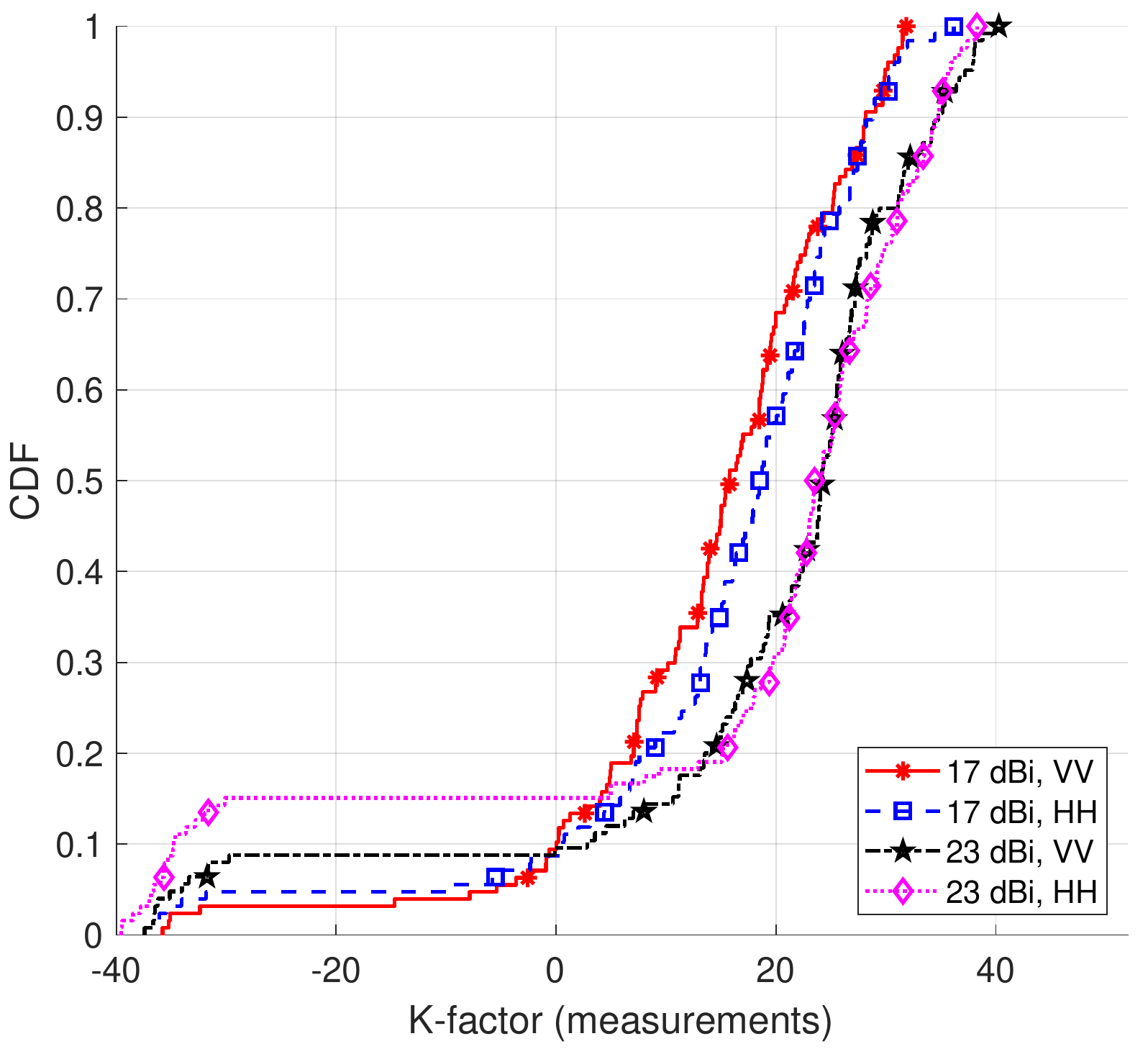}
	  \caption{}  
    \end{subfigure}    
    \begin{subfigure}{\columnwidth}
    \centering
	\includegraphics[width=0.9\textwidth]{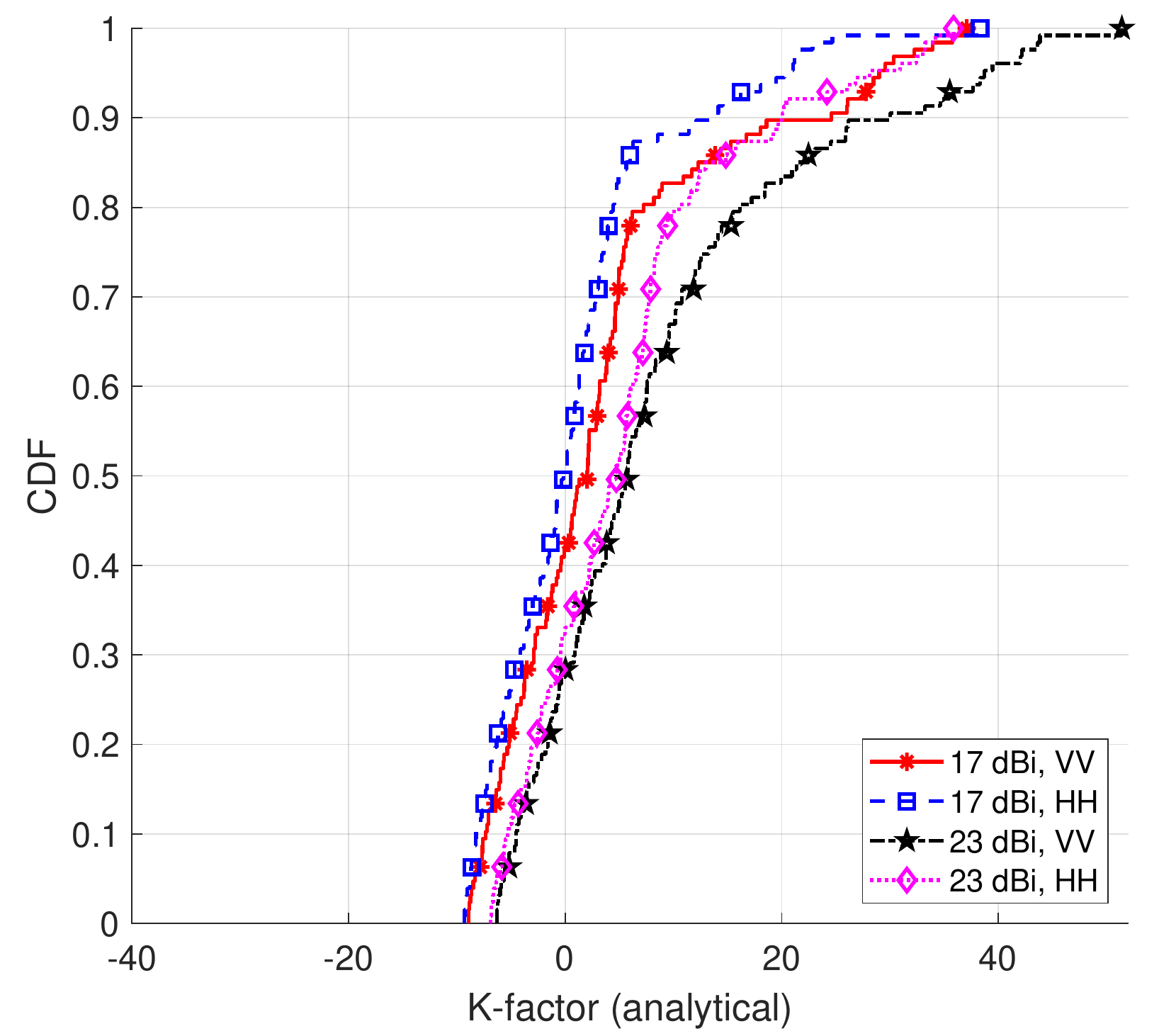}
	  \caption{}  
    \end{subfigure}
     \begin{subfigure}{\columnwidth}
    \centering
	\includegraphics[width=0.9\textwidth]{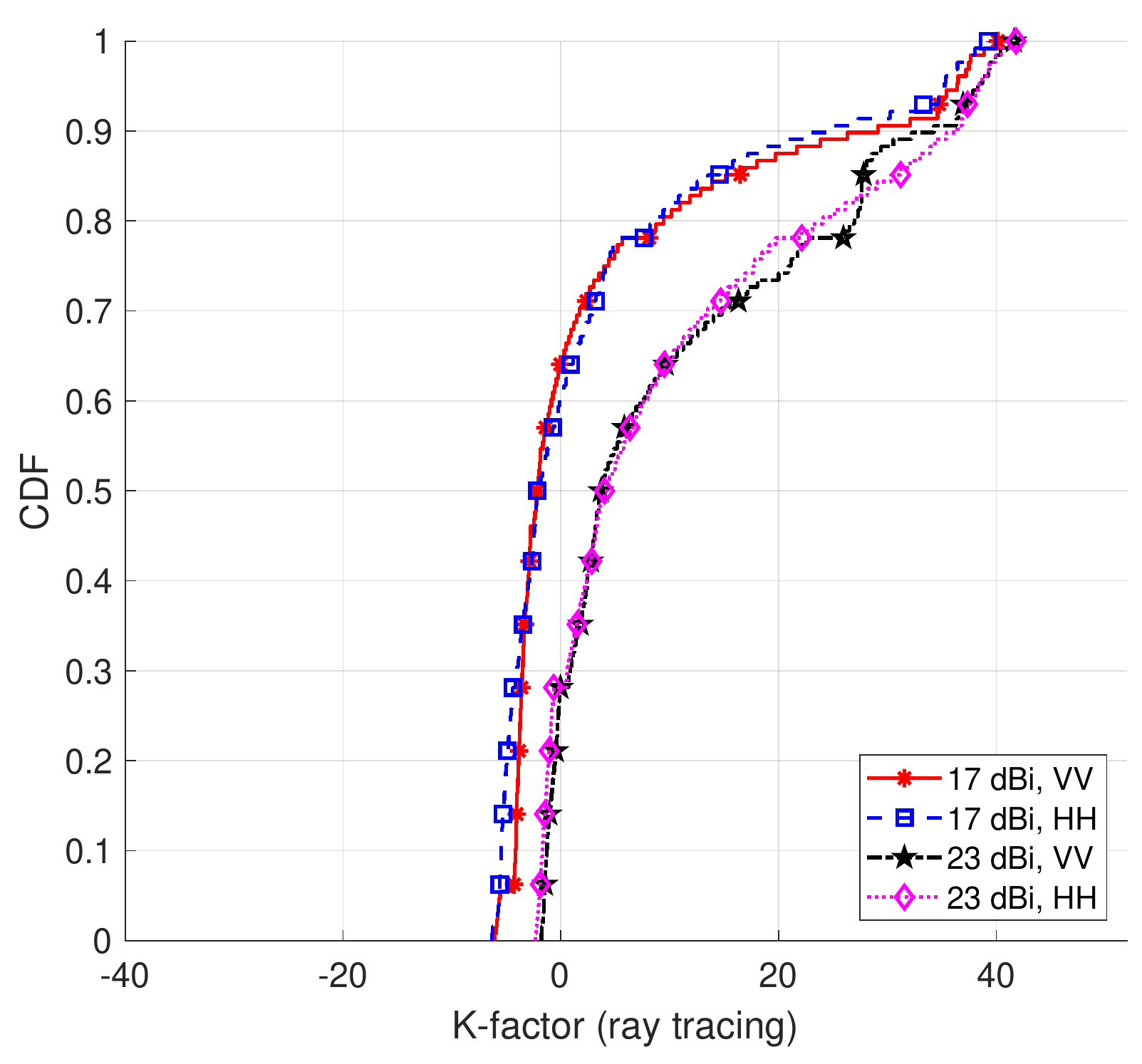}
	  \caption{}  
    \end{subfigure}
    \caption{Ricean $K$-factor for indoor corridor scenario obtained (a) empirically, (b) five ray model~(analytically), (c) ray tracing simulations.}\vspace{-2mm} \label{Fig:K_factor}
\end{figure}
\subsection{Comparison of Five Ray and Two Ray Models}
The two ray and five ray models can be compared with the measurements based on the path loss parameters shown in Table~\ref{Table:Path_loss}. It is observed that line fitting coefficient values~(both slope and y-intercept) of measurement curve and five ray results are close to each other unlike the coefficients of two ray model curve. This is because, taking into account the contribution from additional rays model the propagation better. Moreover, coefficients of the five ray model for $23$~dBi antennas fit measured results better than it fits for $17$~dBi antenna results. Since grazing angles ($\Psi_i$) of second and upper order reflected rays are greater than first order rays grazing angles and since $23$~dBi antennas have narrower beamwidths, it makes sense to have better fit for five ray model results for $23$~dBi antennas.

We can conduct a $z$-test~(as the number of samples is greater than $30$) to better understand the nature of the fittings. The null hypothesis is that the slopes and intercepts are all the same for measurements, and analytical modeling or ray tracing data. Let $\alpha^{\rm (meas)}$, $\alpha^{\rm (anyl)}$ represent the coefficients of slope obtained by linear fitting for measurements and analytical data and $\zeta^{(\alpha_{\rm diff})}$ represents the difference of standard error of the slope fitting; then, the $z$ value is obtained as follows:
\begin{equation}
    z = \frac{\alpha^{\rm (meas)} - \alpha^{\rm (anyl)}}{\zeta^{(\alpha_{\rm diff})}}, \nonumber\
\end{equation}

\begin{equation}
   \zeta^{(\alpha_{\rm diff})}  = \sqrt{\big(\zeta^{\rm (meas)}\big)^2 + \big(\zeta^{\rm (anyl)}\big)^2},  \label{Eq:z_value}
\end{equation}

where $\zeta^{\rm (meas)}$ and $\zeta^{\rm (anyl)}$ are the standard errors of the linear fitted slopes obtained from measurements and analytical modeling data. Similarly, for ray tracing, we have $\alpha^{\rm (RT)}$ and $\zeta^{\rm (RT)}$. A similar expression can be obtained for the y-intercept values.  

The $z$-test values for slope and y-intercept of the linear fittings for analytical and ray tracing compared to the measurements are shown in Table~\ref{Table:z_slope} and Table~\ref{Table:z_intercept}, respectively. It can be observed that we have values of the $z$-test, $1.96<z<-1.96$ for the analytical two ray model for both slope and y-intercept. This indicates that the null hypothesis~(the slope or intercept of measurements is the same as the analytical or ray tracing) can be rejected. On the other hand, for the analytical five ray model, the value of the $z$-test is within the bound [-1.96 1.96] of $95\%$ confidence interval for both slope and y-intercept. This indicates that we have higher chances of the null hypothesis being true. In other words, we can conclude that five ray analytical model provides a better fit for the measurement results compared to the two ray model. Similarly, for ray tracing simulations with five rays, we also have $z$-test values within the bounds of the $95\%$ confidence interval except for the slope fitting of the $17$~dBi antenna at VV antenna orientation.

\section{Ricean $K$-factor} \label{Section:K_factor}
The Ricean $K$-factor is obtained as the ratio of the power of the LOS component to the power of the diffuse components. Fig.~\ref{Fig:K_factor} shows the CDF of the $K$-factor obtained at different link distances indoor for measurements, analytical five ray model and ray tracing simulations. A common observation for all the three is that we have higher $K$-factor for $23$~dBi antennas compared to 17~dBi antennas. This is due to more directional characteristics of the $23$~dBi antenna compared to 17~dBi antenna. The large directivity results in stronger LOS component and weaker diffuse components as observed earlier.

We observe smaller $K$-factor for the measurements compared to analytical and ray tracing results. These small $K$-factor values are due to weak LOS and stronger diffuse components at certain link distances. This is due to destructive interference for the LOS and constructive interference for certain diffuse components. Moreover, we observe larger variance of $K$-factor for $23$~dBi compared to $17$~dBi. This large variance is due to fluctuations in the received power~(as observed in the Section~\ref{Section:RX_power_PL_analysis}).

The $K$-factor for VV and HH antenna orientations are similar for measurement and ray tracing simulations. However, the $K$-factor for VV and HH antenna orientations is different for analytical five ray model. This is mainly due to limited number of paths considered for analytical ray modeling.

\section{Conclusion and Future Work}
\label{Section:Conclusions}
In this paper, we have conducted channel measurements at $28$~GHz in an indoor corridor and outdoor open area. Two horn antenna sets with $17$~dBi and $23$~dBi gains were deployed. Five ray analytical model for received power is compared with two ray model for corridor type indoor environments together with the measured and simulated results taking antenna gains as a hue. Reflections from the ceiling and the side walls increase total received power i.e. the attenuation curve slope in corridor environment is flatter than the free-space path loss curve, and five ray model, which models the environment better than two ray model, gives better results. The attenuation curve gets steeper with increased antenna gain because more directional the antenna means, more rejection of reflected waves approaching from off-boresight directions. Another finding related to this phenomenon is that the difference between received power and the free space path loss increases as the link distance increases, because the grazing angles of the reflected rays get closer to the boresight of the antenna. Another outcome of the work is that in terms of the analytical model curve fitting to measurement results accuracy, curves belonging to $23$~dBi antennas show better performance than $17$~dBi curves because of the increased rejection of higher gain. For the outdoor open area, a two ray model~(a special case of five ray model) is found to provide a better fit to the measurement results than five ray model as expected since because there is no ceiling and sidewalls. In conclusion, the path loss slopes for both indoor corridor and outdoor scenarios were smaller than the free space due to the coherent addition of the reflected rays. Our future work will include modeling of human obstruction for different types of indoor environments e.g. circular tunnels.

\section{Acknowledgements}
This work has been supported in part by NASA under the Federal Award ID number NNX17AJ94A. We also want to thank Kairui Du for his help in the indoor measurements. 

\balance 
\bibliographystyle{IEEEtran}
\bibliography{References}


\end{document}